%% file: PaperGeo.tex
\documentclass[12pt,a4paper]{article}
\usepackage{amssymb,amsfonts,amsmath,amsthm,latexsym,epsfig}

\numberwithin{equation}{section}

\newcommand{\mc}{\mathcal}
\newcommand{\mf}{\mathfrak}

\newcommand{\mat}{\begin{pmatrix}}
\newcommand{\tam}{\end{pmatrix}}
\newcommand{\tr}{\text{tr}}
\newcommand{\rank}{\text{rank}\,}

\newcommand{\eq}{\begin{eqnarray*}}
\newcommand{\qe}{\end{eqnarray*}}
\newcommand{\eqn}{\begin{eqnarray}}
\newcommand{\qen}{\end{eqnarray}}
\newcommand{\Integer}{\mathbb{Z}}
\newcommand{\Real}{\mathbb{R}}
\newcommand{\Complex}{\mathbb{C}}

\newcommand{\bartial}{{\bar{\partial}}}

\def\WZNW{{\text{WZNW}}}
\def\WZ{{\text{WZ}}}
\def\kin{{\text{kin}}}

\def\id{\text{id}}
\def\tr{\text{tr}}
\def\embin{\hookrightarrow}
\def\g{\mathfrak{g}}
\def\h{\mathfrak{h}}
\def\ag{\mathfrak{\hat{g}}}
\def\ah{\mathfrak{\hat{h}}}

\title{On the hierarchy of symmetry breaking\\ D-branes in group manifolds}

\date{\bigskip September 19, 2002}
\author{{\sc Thomas Quella}\footnote{E-mail: quella@aei.mpg.de}\\\\Max-Planck-Institut f\"ur
  Gravitationsphysik\\(Albert-Einstein-Institut)\\Am M\"uhlenberg 1\\D-14476
  Golm\\ Germany}

\begin{document}
  
\baselineskip16pt
\maketitle
\vspace{-10cm}AEI-2002-072\hfill hep-th/0209157\vspace{9cm}

\begin{abstract}
  We construct the boundary WZNW functional for symmetry breaking D-branes
  on a group manifold which are localized along a product of a number of
  twisted conjugacy classes and which preserve an action of an arbitrary
  continuous subgroup. These branes provide a geometric interpretation
  for the algebraic formulation of constructing D-branes developed recently
  in hep-th/0203161. We apply our results to obtain new symmetry breaking
  and non-factorizing D-branes in the background $SL(2,\Real)\times SU(2)$.
\end{abstract}

% -----------------------------------------------------------------------
% -----------------------------------------------------------------------
% -----------------------------------------------------------------------
\section{Introduction}

  The classification of D-branes in given string backgrounds is one of the
  most important tasks in string theory. The interest is mainly
  based on the occurence of non-perturbative dualities between different types
  of string theories and on the natural appearance of (non-commutative) gauge
  theories in the low energy description of open strings.
  The classification of D-branes relies on finding
  conformally invariant boundary conditions for a given two-dimensional bulk
  conformal field theory (CFT). Presently one of the best understood classes
  of CFT's is provided by the Wess-Zumino-Novikov-Witten (WZNW) theories
  \cite{Witten:1984ar,Knizhnik:1984nr} which describe string theory on
  group manifolds \cite{Gepner:1986wi}. Although for dimensional reasons
  most groups cannot be part of a consistent string background, they
  constitute an important ingredient in model building via coset
  constructions \cite{Goddard:1985vk}.
\medskip
  
  In the last few years we have made significant progress with the
  classification of D-branes in a group manifold $G$. For a long time only
  two types have been
  known: untwisted \cite{Cardy:1989ir} and twisted D-branes
  \cite{Birke:1999ik,Behrend:1999bn}.
  Both share the property of preserving the maximal possible symmetry. By now
  we have an almost complete understanding of these D-branes. In particular,
  their geometry was shown to be given by twisted conjugacy
  classes \cite{Alekseev:1998mc,Felder:1999ka}. This observation was supported
  by a Lagrangian description \cite{Gawedzki:1999bq}
  and by a Born-Infeld analysis which also established the stability of these
  D-branes \cite{Bachas:2000ik,Bordalo:2001ec}. Finally, the discovery of the
  noncommutative geometry associated to these branes
  \cite{Alekseev:1999bs,Alekseev:2002rj}
  allowed the discussion of their dynamics
  \cite{Alekseev:2000fd,Fredenhagen:2000ei,Alekseev:2002rj}.
\medskip
  
  In contrast to these successes in the study of maximally symmetric D-branes
  only recently methods have been developed to deal with D-branes which
  break part of the symmetry \cite{Maldacena:2001ky,Maldacena:2001xj,
  Gaberdiel:2001xm,Quella:2002ct}. In the work of Maldacena, Moore and
  Seiberg this was achieved by using a certain kind of T-duality
  related to $U(1)$ subgroups \cite{Maldacena:2001ky,Maldacena:2001xj}.
  A conceptually different and more general algebraic framework which also
  incorporates the usage of non-abelian subgroups was developed in
  \cite{Quella:2002ct}. While the geometric interpretation in the first case
  is obvious from T-duality, a geometric interpretation for more general
  symmetry breaking D-branes was missing until now. In the present work we
  partly fill this gap and show that the most natural symmetry breaking
  D-branes in group manifolds which preserve a given continuous subgroup
  $H\embin G$ are localized along products of quantized twisted conjugacy
  classes of subgroups $U_l$ where
  $H=U_1\embin U_2\cdots\embin U_n=G$. We will prove this fact by constructing
  the corresponding boundary WZNW functional. The identification with D-branes
  obtained from the algebraic description \cite{Quella:2002ct} is 
  established by evaluating the closed string couplings to the branes. It is
  further supported by an analysis of the spectrum of open strings arising from
  both approaches. This comparison relies on the noncommutative geometry
  associated to twisted D-branes \cite{Alekseev:1999bs,Alekseev:2002rj}.
  For completeness we should add that a Lagrangian description of symmetry
  breaking D-branes which are accessible by T-duality has already been found in
  \cite{Sarkissian:2002ie}. These, however, are based on the inclusion
  $U(1)\embin G$ and thus constitute only a small
  subset of an enormous hierarchy of symmetry breaking D-branes arising
  from embedding chains of non-abelian subgroups \cite{Quella:2002ct}.
\medskip
  
  We apply our geometric construction to the classification of
  D-branes in the background $SL(2,\Real)\times SU(2)$. A large number of
  symmetry breaking and non-factorizing D-branes is revealed
  which preserve a continuous subgroup of the target space. All these branes
  may easily be lifted to the string background $AdS_3\times S^3\times T^4$ by
  considering the covering space $AdS_3$ of $SL(2,\Real)$ and using the
  equivalence $SU(2)\cong S^3$. D-branes in the individual factors
  $SL(2,\Real)$ and $SU(2)$ have been described before in \cite{Stanciu:1999nx,
  Bachas:2000fr,Maldacena:2001ky,Rajaraman:2001ew,Sarkissian:2002ie}
  for instance. Our results confirm the geometric description of symmetry
  breaking D-branes in these groups which has been obtained using T-duality.
\medskip

  This paper is organized as follows. In the next section we will construct
  the boundary WZNW functional for symmetry breaking D-branes which are
  localized along the product of twisted conjugacy classes of subgroups $U_l$
  which are organized in an embedding chain
  $H=U_1\embin U_2\cdots\embin U_n=G$. We will argue in section
  \ref{sc:Lagrange} that these D-branes correspond to symmetry breaking
  boundary states which have been constructed in \cite{Quella:2002ct}
  using algebraic methods. Our arguments are based on a target space
  reinterpretation. In the ``new'' target space the D-branes factorize and
  the coupling to closed strings as well as the spectrum of open strings can
  easily be determined. We find full agreement with the Lagrangian approach.
  The general geometric description for symmetry breaking D-branes is
  applied to the classification of D-branes in the group manifold
  $SL(2,\Real)\times SU(2)$ in section \ref{sc:Applications}. Finally, we
  conclude with open problems and further related topics to study.
  
% -----------------------------------------------------------------------
% -----------------------------------------------------------------------
% -----------------------------------------------------------------------
\section{\label{sc:Lagrange}The Lagrangian approach}

  In conformal field theory language, WZNW models are based on a chiral algebra
  $\mc{A}(G)$ which is generated by an affine Kac-Moody algebra $\ag_k$
  at level $k$. The algebra $\ag_k$ is the affine extension of the Lie
  algebra $\g$ belonging to the group manifold $G$, the target space under
  consideration. The bulk symmetry is given by two independent copies
  $\mc{A}(G)\oplus\overline{\mc{A}(G)}$ of the chiral algebra.
  On the boundary holomorphic and antiholomorphic degrees of freedom are
  coupled, and at most one copy of $\mc{A}(G)$ can be preserved. It is natural
  to consider D-branes which preserve only a given subsymmetry
  $\mc{A}(H)\embin\mc{A}(G)$. Here, $\mc{A}(H)$ is the chiral algebra
  generated by an affine Kac-Moody subalgebra $\ah_{k^\prime}\embin\ag_k$. All
  such embeddings arise from continuous subgroups $H$ of $G$.
\medskip
  
  The boundary theory immediately becomes non-rational if one tries to
  preserve only the symmetry $\mc{A}(H)$ on the boundary. Following
  \cite{Quella:2002ct} it is thus natural to consider an embedding chain
  of subgroups\footnote{We will use the phrase embedding map to denote a group
  homomorphism which descends to an injective map on the level of the Lie
  algebras. This condition ensures that the embedding map preserves the
  dimension, but allows for non-trivial wrapping numbers for instance.}
\begin{equation}
  \label{eq:EmbeddingChain}
  H\ =\ U_1\ \embin\ U_2\ \embin\ \cdots\ \embin\ U_{n-1}\ \embin\ U_n
  \ =\ G
\end{equation}
  and to preserve instead the larger chiral algebra
\begin{equation}
  \label{eq:AlgDecompNew}
  \mc{A}(U_1)\ \oplus\ \mc{A}(U_2/U_1)\ \oplus\ \cdots\ \oplus\ \mc{A}(U_n/U_{n-1})\ \embin\ \mc{A}(G)
\end{equation}
  which is rational with respect to the original theory. In writing
  down this expression we denoted by $\mc{A}(U_{l+1}/U_l)$
  the coset chiral algebras which arise via GKO construction
  \cite{Goddard:1985vk}.
  By a straightforward generalization of \cite{Quella:2002ct} we may
  construct D-branes which preserve the chiral algebra
  \eqref{eq:AlgDecompNew} and calculate their spectrum of open strings.
  The geometric interpretation of this kind of D-branes, however, remained
  obscure up to now.  The decomposition \eqref{eq:AlgDecompNew} only implies
  that the submanifolds of $G$ in which these D-branes are localized should
  admit an action of the subgroup $H$.
\medskip

  In this article we provide the missing link between algebra and
  geometry. The remaining part of this section will be devoted to the
  construction of a boundary WZNW functional for D-branes which are localized
  along the product of a number of twisted conjugacy classes, one for each
  subgroup $U_l$. These submanifolds of $G$ admit an action of $H$
  which is inherited from the embedding chain \eqref{eq:EmbeddingChain}. The
  WZNW functional will be shown to be invariant under this action. These
  D-branes thus provide natural candidates for being geometric counterparts of
  those arising from the general algebraic framework \cite{Quella:2002ct}.
  In the next section we will give additional arguments to support this claim.
  This includes the calculation of closed string couplings to the branes
  and an analysis of the spectrum of open strings.
\medskip
  
  Let us be a bit more precise about what we mean by ``product of a number
  of twisted conjugacy classes''. For an exact treatment we have to specify
  embeddings of the individual subgroups $U_l$ into $G$ and an action of
  $H$ on each of these subgroups. Assume that we have given embeddings
  $\epsilon^{U_lU_{l+1}}$ of $U_l$ in $U_{l+1}$ which specify the embedding
  chain \eqref{eq:EmbeddingChain}. In addition we will assume the existence of
  -- possibly trivial -- automorphisms $\Omega_l$ of $U_l$. This
  information may be used to define embeddings
\begin{equation}
  \label{eq:Embeddings}
  \begin{split}
  \epsilon^{HU_l}&\ =\ \epsilon^{U_{l-1}U_l}\circ\cdots\circ\epsilon^{U_2U_3}\circ\epsilon^{U_1U_2}\\
  \epsilon_\Omega^{U_lG}&\ =\ \Omega_n\circ\epsilon^{U_{n-1}U_n}\circ\Omega_{n-1}\circ\cdots\circ\epsilon^{U_{l+1}U_{l+2}}\circ\Omega_{l+1}\circ\epsilon^{U_lU_{l+1}}\ \ .
  \end{split}
\end{equation}
  The embedding $\epsilon_\Omega^{U_lG}$ contains the automorphisms
  $\Omega_k$ at every stage of the embedding chain except for the starting
  point $U_l$. The geometry of our D-branes is the following product of images
  of twisted conjugacy classes,
\begin{equation}
  \label{eq:TwistedConjugacy}
  \mc{D}\bigl\{U_l,\Omega_l,f_l\bigr\}\ =\ \mc{C}_{f_n}^{U_n}\bigl(\Omega_n\bigr)\cdot\epsilon_\Omega^{U_{n-1}G}\Bigl(\mc{C}_{f_{n-1}}^{U_{n-1}}\bigl(\Omega_{n-1}\bigr)\Bigr)\cdot\ \ldots\ \cdot\epsilon_\Omega^{U_1G}\Bigl(\mc{C}_{f_1}^{U_1}\bigl(\Omega_1\bigr)\Bigr)\ \ .
\end{equation}
  The labels $f_l\in U_l$ have to satisfy certain quantization conditions
  and constraints which arise from branching selection rules associated to
  the embedding chain \eqref{eq:EmbeddingChain}. A discussion of these issues
  follows below. The exact definition of the twisted conjugacy classes is
  given by
\begin{equation*}
  \mc{C}_{f_l}^{U_l}\bigl(\Omega_l\bigr)
  \ =\ \bigl\{\ s_l\,f_l\,\Omega_l(s_l^{-1})\ \bigr|\ s_l\in U_l\ \bigr\}\ \ .
\end{equation*}
  Let us denote by $c_l$ elements of the image of twisted conjugacy
  classes in $G$, i.e.\ %
  $c_l\in\epsilon_\Omega^{U_lG}\bigl(\mc{C}_{f_l}^{U_l}\bigl(\Omega_l\bigr)\bigr)$.
  Each of these conjugacy classes admits an action of $H$ which can be
  formulated as follows,
\begin{equation}
  \label{eq:ConjAction}
  s_l\ \mapsto\ \epsilon^{HU_l}(h)\cdot s_l
  \qquad\Rightarrow\qquad
  c_l\ \mapsto\ \epsilon_\Omega^{U_lG}\circ\epsilon^{HU_l}(h)\cdot c_l\cdot\epsilon_\Omega^{U_lG}\circ\Omega_l\circ\epsilon^{HU_l}(h^{-1})\ \ .
\end{equation}
  Due to the recursion relations
  $\epsilon_\Omega^{U_{l+1}G}\circ\Omega_{l+1}\circ\epsilon^{U_lU_{l+1}}=\epsilon_\Omega^{U_lG}$
  also the product of twisted conjugacy classes \eqref{eq:TwistedConjugacy}
  admits a well-defined action of $H$. An arbitrary element
  $x\in\mc{D}\bigl\{U_l,\Omega_l,f_l\bigr\}$ transforms as
\begin{equation}
  \label{eq:AllAction}
  x\ \mapsto \epsilon^{HU_n}(h)\cdot x\cdot\epsilon_\Omega^{U_1G}\circ\Omega_1(h^{-1})\in\mc{D}\bigl\{U_l,\Omega_l,f_l\bigr\}\ \ .
\end{equation}
  As conjectured above, the subset $\mc{D}\bigl\{U_l,\Omega_l,f_l\bigr\}$ of
  $G$ thus indeed provides a natural candidate for a geometric description of
  the symmetry breaking D-branes which arise from the algebraic description via
  the decomposition \eqref{eq:AlgDecompNew}. Notice that our results obviously
  reduce to the usual description of maximally symmetric D-branes by taking
  $\Omega_l=\id$ and $f_l=e$ for $l<n$. It is remarkable that the notion
  of twist becomes much richer for symmetry breaking D-branes in comparison to
  maximally symmetric ones. We are indeed allowed to take non-trivial twists
  for each of the subgroups $U_l$.
\medskip
  
  We are now prepared to substantiate our conjecture by constructing a
  boundary WZNW functional for D-branes localized along the subset
  \eqref{eq:TwistedConjugacy} which is invariant under the action
  \eqref{eq:ConjAction} of $H$.
  Geometrically, the theory is described by a non-linear $\sigma$-model
  of fields $g:\Sigma\to G$ which live on a two-dimensional world sheet
  $\Sigma$ and which take values in the group manifold $G$. We will always
  assume that $G$ is simple in what follows, but it is straightforward to
  generalize our results to reductive groups, i.e. to those which are a direct
  product of simple groups and $U(1)$ factors. The action functional for this
  theory is given by
\begin{equation}
  \label{eq:BoundaryAction}
  S_{\WZNW}^{G}\bigl(g;k|\mc{D}\bigr)
  \ =\ S_{\kin}^{G}(g;k)\ + \ S_{\WZ}^{G}(g;k)\ +\ S_{\mc{D}}^{G}(g;k)
\end{equation}
  and consists of three parts, the usual kinetic term, the so-called
  Wess-Zumino term and a boundary term. All of them contain a parameter $k>0$
  subject to certain consistency conditions (see below). The kinetic term is
  given by
\begin{equation*}
  S_{\kin}^{G}(g;k)
  \ =\ -\frac{k}{4\pi}\frac{2}{I_R}\int_\Sigma d^2z\:\tr_R\bigl\{\partial gg^{-1}\bartial gg^{-1}\bigr\}\ \ .
\end{equation*}
  The trace is evaluated in some non-trivial unitary representation $R$ of the
  Lie algebra $\mf{g}$ of $G$. This is indicated by the explicit appearance of
  the Dynkin index $I_R$ of the representation. The symbol $\tr_R$ denotes the
  trace of $\dim R$-dimensional matrices. The combination $2/I_R\:\tr_R$ is a
  normalized trace which is independent of the representation $R$. In our
  conventions the Killing form is obtained from
  $I_R\,\kappa^{\alpha\beta}=\tr_R\bigl\{R(T^\alpha)R(T^\beta)\bigr\}$.
\medskip

  The Wess-Zumino term is defined in terms of its associated $3$-form 
  $\omega^{\text{WZ}}$. Its contribution to the boundary WZNW functional
  \eqref{eq:BoundaryAction} is given by
\begin{equation*}
  S_{\WZ}^{G}(g;k)
  \ =\ -\frac{k}{4\pi}\frac{2}{I_R}\int_B\ \omega^{\text{WZ}}
  \qquad\text{ with }\qquad
  \omega^{\text{WZ}}(g)\ =\ \frac{1}{3}\,\tr_R(g^{-1}dg)^3\ \ .
\end{equation*}
  This integral extends over a three-dimensional manifold $B$ whose boundary
  is given by $\partial B=\Sigma\cup D$ where $D$ is a disjoint union of
  (topological) discs filling the holes of $\Sigma$ such that $\Sigma\cup D$
  has no boundaries. For notational simplicity we will assume that $D$
  consists of exactly one disc.
\medskip
  
  If one only considers the first two terms of the action functional
  \eqref{eq:BoundaryAction}, it is invariant under all transformations
  $g\mapsto g_L(z)\,g\,g_R^{-1}(\bar{z})$. This is the famous loop group
  symmetry $G(z)\times G(\bar{z})$ of the WZNW theory. This symmetry is
  generated by two Lie algebra valued currents $J(z)=-k\partial g g^{-1}$ and
  $\bar{J}(\bar{z})=kg^{-1}\bartial g$ which are chiral by the equations of
  motion. They generate two commuting copies of the affine Kac-Moody algebra
  $\mf{\hat{g}}_k$ at level $k$ as mentioned already in the beginning of
  this section.
\medskip
  
  To complete the definition of the boundary WZNW functional
  \eqref{eq:BoundaryAction} we finally have to define the boundary term.
  This is given by the integral
\begin{equation}
  \label{eq:BoundaryTerm}
  S_{\mc{D}}^{G}(g;k)
  \ =\ \frac{k}{4\pi}\frac{2}{I_R}\int_D\ \omega_{\mc{D}}
  \ =\ \frac{k}{4\pi}\frac{2}{I_R}\int_D\ \sum_{l=1}^n\sum_{k=1}^l\:\omega_{\mc{D}}(c_k,\cdots,c_l)
\end{equation}
  over the auxiliary disc $D$. We assume that the boundary of $\Sigma$ and the
  whole disc $D$ are mapped into the subset
  $\mc{D}\bigl\{U_l,\Omega_l,f_l\bigr\}$ of $G$. The first condition justifies
  the use of the word D-brane when referring to this submanifold. The boundary
  two-forms entering the definition \eqref{eq:BoundaryTerm} are specified by
\begin{equation}
  \label{eq:BoundaryForm}
  \begin{split}
  \omega_{\mc{D}}(c_l)
  &\ =\ \tr_R\Bigl\{\epsilon^{U_lG}\Bigl(s_l^{-1}ds_l\,f_l\,\Omega_l\bigl(s_l^{-1}ds_l\bigr)\,f_l^{-1}\Bigr)\Bigr\}\\
  \omega_{\mc{D}}(c_k,\cdots,c_l)
  &\ =\ -\tr_R\bigl\{c_k^{-1}\cdots c_l^{-1}dc_lc_{l-1}\cdots c_{k+1}dc_k\bigr\}\ \ .
  \end{split}
\end{equation}
  If $\mc{D}\bigl\{U_l,\Omega_l,f_l\bigr\}$ is a single twisted conjugacy
  class, the expression \eqref{eq:BoundaryTerm} reduces to those found for
  maximally symmetric D-branes \cite{Gawedzki:1999bq}. For a product of two
  twisted conjugacy classes we recover boundary terms which have been used to
  describe maximally symmetric D-branes in coset spaces
  \cite{Gawedzki:2001ye,Elitzur:2001qd}.
  Our expressions also contain as a special case the recent results of
  \cite{Sarkissian:2002ie} by taking the ``embedding chain'' $U(1)\embin G$
  and a particular choice of automorphisms. The exact correspondence will be
  subject of section \ref{sc:Applications}.
\medskip
  
  After having provided the complete definition of the boundary WZNW functional
  a few remarks are in order. The physics which is described by the action
  \eqref{eq:BoundaryAction} should not depend on the two auxiliary
  manifolds $B$ and $D$. This leads to restrictions such as quantization
  conditions for the level and allowed conjugacy classes as well as
  branching selection rules
  \cite{Gawedzki:1999bq,Gawedzki:2001ye,Elitzur:2001qd}. Let us be
  a little bit more specific. For compact simply-connected simple Lie groups
  topological considerations regarding the Wess-Zumino term force $k$ to be a
  positive integer. This may be different for non-simply-connected or
  non-compact groups. In the case of $G=SO(3)$ the level $k$ has to be even
  for instance and for $G=SL(2,\Real)$ we obtain no additional constraints on
  the level. Invariance of the action functional \eqref{eq:BoundaryAction}
  under infinitesimal deformations of the disc is ensured by the relation
  $d\omega_{\mc{D}}=\omega^{\text{WZ}}\bigr|_{\mc{D}}$ which is proven
  in the appendix. Taking global aspects of the embedding of the disc into
  account one exactly recovers the quantization of twisted conjugacy classes
  and branching selection rules which would be expected from the CFT
  description \cite{Gawedzki:1999bq,Gawedzki:2001ye,Elitzur:2001qd}.
  This concludes our arguments that the action \eqref{eq:BoundaryAction} is
  well-defined.
\medskip

  As was discussed in \cite{Gawedzki:2001ye,Elitzur:2001qd} the boundary
  two-form $\omega_{\mc{D}}$ does not only depend on the values of the field
  $g$ on the boundary but also on the exact decomposition into a product
  $g=c_n\cdots c_1$ of elements of the individual twisted conjugacy classes.
  Under certain circumstances the sets $\mc{D}\bigl\{U_l,\Omega_l,f_l\bigr\}$
  and $\mc{D}\bigl\{U_l^\prime,\Omega_l^\prime,f_l^\prime\bigr\}$ are identical.
  The algebraic analysis, however, suggests that they should describe different
  D-branes with different
  spectrum and different mass density. This becomes particularly important for
  products of twisted conjugacy classes which cover the whole group $G$.
  The solution to this puzzle is presented in section \ref{sc:TargetSpace} where
  it is shown that it is natural to adopt a different target space
  interpretation when using the decomposition \eqref{eq:AlgDecompNew}.
  In this larger target space the shape of all our D-branes is distinct.
\medskip

  It remains to be shown that the boundary WZNW functional
  \eqref{eq:BoundaryAction} is invariant under the action
  (\ref{eq:ConjAction},\,\ref{eq:AllAction})
  of $H$ on the boundary. To be specific, the symmetry $G(z)\times G(\bar{z})$
  of the bulk theory has
  to be broken to a symmetry $H(\tau)\embin G(\tau)\times G(\tau)$ on the
  boundary where the embedding is given by
  $\bigl(\epsilon^{HU_n},\epsilon_\Omega^{U_1G}\circ\Omega_1\bigr)$.
  With the same embedding map one can define an action of
  $H(z)\times H(\bar{z})$ in the bulk theory which induces the decomposition
  \eqref{eq:AlgDecompNew} of chiral algebras. This reduced action gives rise
  to chiral currents $J^{\h}(z)$ and $\bar{J}^{\h}(\bar{z})$ which take
  values in the Lie algebra $\mf{h}$. Demanding that this action reduces to
  the action of $H(\tau)$ on the boundary is equivalent to enforcing trivial
  gluing conditions on the currents $J^{\h}(z)$ and
  $\bar{J}^{\h}(\bar{z})$. Note that it is not relevant whether one demands
  twisted gluing conditions or puts the twist in the definition of the
  current as the decomposition of the bulk Hilbert space also has to reflect
  this choice. In our previous considerations we adopted the second point of
  view.
\medskip

  Let us now determine the variation of the boundary WZNW functional under an
  arbitrary infinitesimal action of $h(\tau)=1+i\omega(\tau)\in H(\tau)$.
  A lengthy but straightforward calculation results in
\begin{equation*}
  \delta\omega_{\mc{D}}
  \ =\ -i\,\sum_{l=1}^n\tr\bigl\{d\omega_L^{(n)}c_n\cdots c_{l+1}dc_lc_l^{-1}\cdots c_n^{-1}+d\omega_R^{(1)}c_1^{-1}\cdots c_l^{-1}dc_lc_{l-1}\cdots c_1\bigr\}\ \ ,
\end{equation*}
  where we introduced the abbreviations 
  $\omega_L^{(n)}=\epsilon^{HU_n}(\omega)$
  and $\omega_R^{(1)}=\epsilon_\Omega^{U_1G}\circ\Omega_1(\omega)$.
  The variation of the Wess-Zumino term may be determined from
\begin{equation}
  \label{eq:VariationWZ}
  \delta \omega^{\text{WZ}}
  \ =\ -i\,d\,\tr\Bigl\{d\omega_L^{(n)}dgg^{-1}+d\omega_R^{(1)}g^{-1}dg\Bigr\}\ \ .
\end{equation}    
  After integration it will give two contributions which arise from the
  boundary $\Sigma\cup D$ of $B$. The first one belonging to $\Sigma$ is
  canceled by the variation of the kinetic term. If we restrict the discussion
  to the disc which is mapped to the set
  $\mc{D}\bigl\{U_l,\Omega_l,f_l\bigr\}$, the variation \eqref{eq:VariationWZ}
  further simplifies to
\begin{equation*}
  \delta \omega^{\text{WZ}}\bigr|_{D}
  \ =\ -i\,\sum_{l=1}^nd\,\tr\Bigl\{d\omega_L^{(n)}c_n\cdots c_{l+1}dc_lc_l^{-1}\cdots c_n^{-1}+d\omega_R^{(1)}c_1^{-1}\cdots c_l^{-1}dc_lc_{l-1}\cdots c_1\Bigr\}\ \ .
\end{equation*}
  Obviously, the contributions from $\delta\omega^{\text{WZ}}$ and
  $\delta\omega_{\mc{D}}$ cancel each other exactly. The details of the
  calculation can be found in the appendix. This completes the proof
  of the symmetry of the D-branes $\mc{D}\bigl\{U_l,\Omega_l,f_l\bigr\}$ under
  the action of the subgroup $H$.
\medskip

  The procedure described in this section provides a whole hierarchy of
  symmetry breaking D-branes. The classification of all these objects is
  greatly simplified by the following observation which is well-known
  from maximally symmetric D-branes. Instead of allowing all choices
  of automorphisms we may restrict to the case of outer automorphisms.
  The appearance of an inner automorphism $\Omega_l(u_l)=b_lu_lb_l^{-1}$
  in the product of twisted conjugacy classes \eqref{eq:TwistedConjugacy}
  just corresponds to putting this automorphism to the identity
  in the expressions \eqref{eq:Embeddings} and \eqref{eq:TwistedConjugacy},
  to replace $f_l$ by $f_lb_l$ and to multiply the resulting expression for
  eq.\ \eqref{eq:TwistedConjugacy} with the
  element $\epsilon_\Omega^{U_lG}(b_l^{-1})$ from the right. Geometrically,
  this procedure induces an overall shift. The same idea also enables us
  to choose specific representatives for whole classes of outer
  automorphisms by separating their ``inner part''. 
\medskip

  One may ask whether arranging the twisted conjugacy classes in the
  definition \eqref{eq:TwistedConjugacy} in different order would lead to
  new results. It is easy to show that this is not the case. Exchanging two
  conjugacy classes merely leads to a redefinition of embedding maps and
  automorphisms. Under these circumstances it is natural to work with
  one standard representative. In our case the latter is defined to be
  given by eq.\ \eqref{eq:TwistedConjugacy}.
\medskip

  We conclude with a few remarks. The first
  concerns the dimension of the D-branes which correspond to
  the product of twisted conjugacy classes \eqref{eq:TwistedConjugacy}.
  For a naive evaluation of the dimension one would simply add the dimensions
  of the twisted conjugacy classes present in eq.\ %
  \eqref{eq:TwistedConjugacy}. It is obvious that this procedure would rapidly
  exceed the dimension of the
  group itself if one takes embedding chains \eqref{eq:EmbeddingChain}
  with a large number of subgroups. Up to now we lack a general dimension
  formula for this kind of D-branes. Let us emphasize at this point the
  remarkable fact that they tend to be more and more space-filling the more of
  the symmetry we break. We believe that branes which cover the whole target
  space can be constructed for every WZNW model. A natural candidate for such
  a space-filling
  brane is obtained by taking the product of a non-degenerate ordinary
  conjugacy class of $G$ and a distinguished twisted conjugacy class of its
  maximal torus $T=U(1)^{\rank G}$. The first is isomorphic to $G/T$
  \cite{Felder:1999ka}, i.e.\ has dimension $\dim G-\rank G$, while the second
  is given by $T$ itself. In section \ref{sc:Applications} we will confirm
  our conjecture in a number of examples.
\medskip

  The second remark concerns the hierarchical structure of the symmetry
  breaking D-branes which are described by the product of twisted conjugacy
  classes \eqref{eq:TwistedConjugacy}. Let us consider a fixed embedding
  chain \eqref{eq:EmbeddingChain}. By choosing an automorphism $\Omega_l$
  to act trivially and the corresponding element $f_l$ to be given by the group
  unit we can achieve that the conjugacy class
  $\mc{C}_{f_l}^{U_l}\bigl(\Omega_l\bigr)$ may be omitted from the
  expression \eqref{eq:TwistedConjugacy} for the geometry of the D-brane.
  This means that we could have equally well omitted the group $U_l$ from the
  embedding chain \eqref{eq:EmbeddingChain} in order to describe the same
  D-brane. The same feature has also been observed in the algebraic description
  \cite{Quella:2002ct}. To obtain a classification of D-branes in the group
  manifold $G$ which preserve an arbitrary continuous subgroup it is thus
  enough to find all inequivalent chains of maximal embeddings.

% -----------------------------------------------------------------------
% -----------------------------------------------------------------------
% -----------------------------------------------------------------------
\section{\label{sc:Algebra}The algebraic point of view}

  In the last section we used the Lagrangian approach to construct a large
  number of D-branes on a group manifold $G$ which preserve a given continuous
  subgroup $H$. We also presented first indications that these
  provide the geometric interpretation for symmetry breaking boundary states
  arising in the algebraic approach \cite{Quella:2002ct}. In the following
  we will further illuminate this equivalence. Full agreement is found
  when calculating the coupling of closed strings to the branes. Also the
  spectrum of open strings which is predicted by geometry fits into the
  algebraic description. Before we dive into the discussion of branes we
  present a natural target space reinterpretation which sheds new light on some
  open issues which have been the cause of some confusion in the past.
  As no general tools are available for an algebraic description of
  non-compact WZNW theories, we assume the group $G$ to be compact in
  what follows. To simplify notation the group is also considered to be
  simply-connected and simple.

% -----------------------------------------------------------------------
% -----------------------------------------------------------------------
\subsection{\label{sc:TargetSpace}Target space reinterpretation}

  The usual interpretation of a WZNW theory relies on the group $G$ itself
  as target space. In the context of the decomposition \eqref{eq:AlgDecompNew}
  of the chiral algebra it is, however, more convenient to work with the
  space
\begin{equation}
  \label{eq:Extension}
  G^{\text{new}}\ =\ \frac{U_n\times U_{n-1}\times U_{n-1}\times\cdots\times U_1\times U_1}{U_{n-1}\times U_{n-1}\times\cdots\times U_1\times U_1}
  \ =\ \frac{G\times X}{X}\ \ .
\end{equation}
  This statement is a generalization of a proposal which has been formulated
  in the context of coset theories \cite{Fredenhagen:2001kw}. The specific
  form of the auxiliary space $X$ has its origin in the decomposition of
  chiral algebras \eqref{eq:AlgDecompNew}. It is motivated by the deep
  relation between coset CFT's $U_{l+1}/U_l$ and product CFT's
  $U_{l+1}\times U_l$ which itself is based on the similarity of modular
  properties. The extension of $G$ by $X$ introduces additional degrees of
  freedom which have to be removed by dividing through $X$. The exact
  action of $X$ will be given in eq.\ \eqref{eq:Identification}.
  We will argue below that the product of twisted conjugacy classes
  \eqref{eq:TwistedConjugacy} which has been defined on $G$ possesses a natural
  interpretation as a direct product in the numerator $G\times X$
  of the new target space $G^{\text{new}}$. In this picture each twisted
  conjugacy class of a group $U_l$ with $l<n$ is diagonally embedded in the
  product $U_l\times U_l$.
\medskip
  
  The equivalence of the spaces $G$ and $G^{\text{new}}$ seems to be obvious
  at first sight. Nevertheless we have to be very careful as $G$ carries
  additional structure which should be reflected in $G^{\text{new}}$.
  In particular, $G$ admits an action of the group $G\times G$, i.e.\ the
  regular action from the left and from the right. The group $G\times G$
  should be considered as the ``constant'' part of the WZNW symmetry
  $G(z)\times G(\bar{z})$ which has been described in the last section.
  When we consider symmetry breaking D-branes which arise from the embedding
  chain \eqref{eq:EmbeddingChain}, the action of $G\times G$ thus has to be
  broken to an action of the subgroup $H\times H$ where the embedding of the
  latter is given by the map
  $\bigl(\epsilon^{HU_n},\epsilon_\Omega^{U_1G}\circ\Omega_1\bigr)$. We will
  argue below that the same action of $H\times H$ can be found on
  $G^{\text{new}}$ provided that one uses the correct action of $X$ on
  $G\times X$ in the definition \eqref{eq:Extension}. To be precise, the
  elements of $G^{\text{new}}$ should be given by tupels
  $(u_n,u_{n-1}^\prime,u_{n-1},\cdots,u_2^\prime,u_2,u_1^\prime,u_1)\in G\times X$
  subject to the identifications
\begin{align}
  \label{eq:Identification}
  (u_l^\prime,u_l)&\ \sim\ (u_l^\prime\cdot t_l^{-1},t_l\cdot u_l)&
  &\text{and}&
  (u_{l+1},u_l^\prime)&\ \sim\ (u_{l+1}\cdot\Omega_{l+1}\circ\epsilon^{U_lU_{l+1}}(s_l^{-1}),s_l\cdot u_l^\prime)
\end{align}
  for $t_l,s_l\in U_l$. The action of $H\times H$ on the target space
  $G^{\text{new}}$ on the other hand should be defined by
  $(u_n,\cdots,u_1)\mapsto\bigl(\epsilon^{HU_n}(h_1)u_n,\cdots,u_1\Omega_1(h_2^{-1})\bigr)$. The identification \eqref{eq:Identification} shows that the new
  target space $G^{\text{new}}$ is a specific example of an asymmetric coset.
  A general discussion of string theory in asymmetric coset spaces will
  follow in \cite{TQVS:Unpublished}.
\medskip
  
  The connection of the new target space $G^{\text{new}}$ to the
  considerations in the previous section may easily be illustrated by working
  out the natural representatives of elements in $G^{\text{new}}$. They are
  given by
\begin{equation}
  \label{eq:Representative}
  \bigl(u_n\cdot\epsilon_\Omega^{U_{n-1}G}(u_{n-1}^\prime\cdot u_{n-1})\cdot\ldots\cdot\epsilon_\Omega^{U_1G}(u_1^\prime\cdot u_1),e,\ldots,e\bigr)\ \ .
\end{equation}
  A comparison with eq.\ \eqref{eq:TwistedConjugacy} indicates that the
  product of twisted conjugacy classes which appeared in the last section
  has a natural interpretation in the new target space $G^{\text{new}}$.
  The relation \eqref{eq:Representative} indeed shows that elements of
  $G^{\text{new}}$ may be represented naturally as elements of $G$. Note
  however that one and the same element of $G$ is represented by a whole
  orbit of elements in $G\times X$. This makes explicit the drastical
  increase of degrees of freedom which are associated to the decomposition
  \eqref{eq:AlgDecompNew} of the chiral algebra. What happened to be a 
  cause of confusion in \cite{Gawedzki:2001ye,Elitzur:2001qd} and in the
  previous section, has now found its natural explanation. On the contrary,
  it seems to give us the possibility to describe new interesting features
  such as superpositions of D-branes and multiple wrappings
  (see also \cite{Maldacena:2001ky}).
\medskip

  The reader may wonder why we had to choose such a complicated identification
  \eqref{eq:Identification} to define the coset $G^{\text{new}}$. A partial
  answer was given already by the striking relation between the representative
  \eqref{eq:Representative} and the form of the product of twisted conjugacy
  classes \eqref{eq:TwistedConjugacy}. The deeper reason for this particular
  choice of identification comes, however, from demanding the equality of
  $G$ and $G^{\text{new}}$ including the given action of $H\times H$ on them.
  The latter can be understood best if one does not work on the level of
  manifolds but descends to the algebras of functions $\mc{F}(G)$ and
  $\mc{F}(G^{\text{new}})$ which inherit the given action of $H\times H$
  but allow for a {\em linear} representation. According to a theorem of
  Gel'fand and Naimark also the topology of a manifold is completely contained
  in its algebra of functions. Showing the equality
\begin{equation}
  \label{eq:Isomorphy}
  \mc{F}(G)\ \cong\ \mc{F}(G^{\text{new}})\ =\ \text{Inv}_{X}\Bigl(\mc{F}\bigl(G\times X\bigr)\Bigr)
\end{equation}
  as $H\times H$ modules is thus enough to establish
  the equivalence of the target spaces $G$ and $G^{\text{new}}$ including the
  action of $H\times H$ on them. The relation \eqref{eq:Isomorphy} may be
  proven by using the Peter-Weyl theorem which gives the decomposition of the
  algebra of functions on a group into irreducible representations under left
  and right regular action of the group itself. Restricting the action to
  $H\times H$ and taking all twists into account we exactly recover the
  equality \eqref{eq:Isomorphy}.
\medskip

  The considerations of the last paragraph have direct implications for the
  conformal field theory description. Let us bring to mind that the
  interpretation of $G$ as the target space of a WZNW theory is
  supported by the deep relation between the spectrum of closed strings and
  the algebra of functions on the group. According to the Peter-Weyl theorem
  the algebra of functions $\mc{F}(G)$ is recovered from the ground state
  structure of the charge conjugate partition function of the WZNW theory in
  the limit $k\to\infty$ when interpreted as a $G\times G$ module with respect
  to left and right regular action of $G$. As already mentioned above, the
  group $G\times G$ should be considered as the ``constant'' part of
  $G(z)\times G(\bar{z})$. After symmetry reduction, the decomposition
  \eqref{eq:AlgDecompNew} of the chiral algebras has to be accompanied by an
  analogous decomposition of the closed string Hilbert space. On the
  geometrical side this corresponds to the interpretation of the $G\times G$
  module $\mc{F}(G)$ as an $H\times H$ module where the embedding is given by
  $\bigl(\epsilon^{HU_n},\epsilon_\Omega^{U_1G}\circ\Omega_1\bigr)$.

% -----------------------------------------------------------------------
% -----------------------------------------------------------------------
\subsection{The coupling of boundary states to closed strings}

  The main aim of this section is to understand the relation between
  the geometric results of section \ref{sc:Lagrange} and the algebraic method
  of constructing symmetry breaking boundary states \cite{Quella:2002ct}.
  It is thus necessary to recapitulate the main formulas of this approach.
  We will be rather sketchy in what follows. In particular we will not be
  concerned with technical difficulties such as field identification or
  related topics. We will tacitly assume that this phenomenon does not appear
  for the cases under consideration. In the geometric regime where the level
  $k$ runs to infinity, this seems to be a valid approximation in most of the
  cases. The interested reader is referred to \cite{Quella:2002ct} where he can
  find the missing details. The general framework of boundary conformal field
  theory is subject of \cite{Recknagel:1998sb,Fuchs:1998fu,Fuchs:1999zi,
  Fuchs:1999xn,Behrend:1999bn,Fuchs:2002cm} for instance.
\medskip

  The decomposition of the chiral algebras \eqref{eq:AlgDecompNew} is
  accompanied by a decomposition
\begin{equation}
  \mc{H}_{\mu_n}^G\ =\ \bigoplus \mc{H}_{(\mu_n,\mu_{n-1})}^{U_n/U_{n-1}}\otimes\cdots\otimes\mc{H}_{(\mu_2,\mu_1)}^{U_2/U_1}\otimes\mc{H}_{\mu_1}^{U_1}
\end{equation}
  of representation spaces $\mc{H}_{\mu_n}^G$ of $\mc{A}(G)$. It is very
  important, however, that the decomposition of the antiholomorphic part looks
  slightly different as it has to reflect the different choices of $H$-actions
  which are used to define the currents $J^{\h}(z)$ and
  $\bar{J}^{\h}(\bar{z})$. To be precise we obtain
\begin{equation}
  \mc{\bar{H}}_{\mu_n}^G\ =\ \bigoplus \mc{\bar{H}}_{(\Omega_n^{-1}(\mu_n),\nu_{n-1})}^{U_n/U_{n-1}}\otimes\cdots\otimes\mc{\bar{H}}_{(\Omega_2^{-1}(\nu_2),\nu_1)}^{U_2/U_1}\otimes\mc{\bar{H}}_{\Omega_1^{-1}(\nu_1)}^{U_1}\ \ .
\end{equation}
  We observe that the weights $\nu_l$ appear with a relative twist in the
  different representation spaces. The last two relations induce an analogous
  decomposition of the full charge conjugate closed string Hilbert space
  $\mc{H}^G$.
\medskip

  Before we are able to construct the boundary states which contain all the
  relevant data to perform the comparison with the geometric picture we need
  some further preparations. Assume that we have given the algebraic solution
  corresponding to twisted gluing conditions in the auxiliary algebras
  $\mc{A}(U_l)$. This means that we know structure constants $\psi^{U_l}$ such
  that the matrices
\begin{equation}
  \label{eq:NIMrep}
  \Bigl(n_{\nu_l}^{U_l}\Bigr)_{\rho_l}^{\rho_l^\prime}
  \ =\ \sum_{\Omega_l(\mu_l)=\mu_l}\frac{\bigl(\bar{\psi}^{U_l})_{\rho_l^\prime}^{\mu_l}\bigl(\psi^{U_l}\bigr)_{\rho_l}^{\mu_l}S_{\mu_l\nu_l}^{U_l}}{S_{\mu_l0}^{U_l}}
\end{equation}
  form a NIM-rep of the fusion algebra of the affine Kac-Moody algebra
  $(\mf{\hat{u}}_l)_{k_l}$. In this expression we denoted by $S^{U_l}$ the
  modular S matrix of $(\mf{\hat{u}}_l)_{k_l}$. The structure constants
  $\bigl(\psi^{U_l}\bigr)_{\rho_l}^{\mu_l}$ carry two labels. One of them
  refers to boundary conditions $\rho_l$ while the other refers to symmetric
  representations $\mu_l=\Omega_l(\mu_l)$ of
  $(\mf{\hat{u}}_l)_{k_l}$. The simplest example of a NIM-rep may be obtained
  from $\Omega_l=\id$ by putting
  $\bigl(\psi^{U_l}\bigr)_{\rho_l}^{\mu_l}=S_{\mu_l\rho_l}^{U_l}$. In this
  case, Cardy's case, the matrices $n_{\nu_l}^{U_l}$ are just the fusion
  matrices. Details of these constructions can be found
  in \cite{Birke:1999ik,Fuchs:1999zi,Behrend:1999bn} for instance.
\medskip

  Imposing trivial gluing conditions on all constituents of the reduced
  chiral algebra \eqref{eq:AlgDecompNew} enforces the condition
  $\mu_l=\nu_l=\Omega_l(\nu_l)$ for the symmetric part of $\mc{H}^G$ from
  which we may construct Ishibashi states
  $|\mu_n,\cdots,\mu_1\rangle\!\rangle$ \cite{Quella:2002ct}. By comparison
  with the previous paragraph we are thus able to define boundary states
\begin{equation}
  \label{eq:BoundaryState}
  |\rho_n,\cdots,\rho_1\rangle\ =\ \sum\:\frac{\bigl(\psi^{U_n}\bigr)_{\rho_n}^{\mu_n}}{\sqrt{S_{0\mu_n}^{U_n}}}\,\frac{\bigl(\psi^{U_{n-1}}\bigr)_{\rho_{n-1}}^{\mu_{n-1}}}{S_{0\mu_{n-1}}^{U_{n-1}}}\,\cdots\,\frac{\bigl(\psi^{U_2}\bigr)_{\rho_2}^{\mu_2}}{S_{0\mu_2}^{U_2}}\,\frac{\bigl(\psi^{U_1}\bigr)_{\rho_1}^{\mu_1}}{S_{0\mu_1}^{U_1}}\:|\mu_n,\cdots,\mu_1\rangle\!\rangle\ \ .
\end{equation}
  Up to this point we merely reviewed the algebraic method of constructing
  symmetry breaking boundary states \cite{Quella:2002ct}. We will now use
  this algebraic information to prove that these boundary states describe
  D-branes which are localized along the product of twisted conjugacy classes
  \eqref{eq:TwistedConjugacy}. The first argument is based on the coupling
  of closed strings to the brane. In the section \ref{sc:Spectrum} we will
  then analyze the spectrum of open strings.
\medskip
  
  It is convenient to think about the boundary state \eqref{eq:BoundaryState}
  as describing D-branes in the group manifold
  $U_n\times U_{n-1}\times U_{n-1}\times\cdots\times U_1\times U_1$
  which is the basic constituent of $G^{\text{new}}$. This follows from the
  discussion in section \ref{sc:TargetSpace} and the factorized
  structure of eq.\ \eqref{eq:BoundaryState}. A similar proposal
  in the context of coset theories has been made in \cite{Fredenhagen:2001kw}
  (see also \cite{Ishikawa:2002wx}).
  According to the general discussion of \cite{Felder:1999ka}, the calculation
  of closed string couplings to the brane shows that the $U_n$-part of the
  boundary state \eqref{eq:BoundaryState} gives rise to a D-brane which is
  localized along the twisted conjugacy class in
  $\mc{C}_{f_n}^{U_n}\bigl(\Omega_n\bigr)$ in $U_n$. The element $f_n$ is
  obtained from exponentiation of the weight $\rho_n$ to the (symmetric)
  Cartan torus in $U_n$. This prescription also makes explicit the
  quantization conditions which have to be imposed on twisted conjugacy
  classes. The other parts in eq.\ \eqref{eq:BoundaryState}
  are more difficult to access as the coefficients possess an additional
  factor of $\bigl(S_{0\mu_l}^{U_l}\bigr)^{1/2}$ in the denominator. One may,
  however, easily check that such coefficients appear when constructing
  D-branes in $U_l\times U_l$ if one uses the gluing automorphism
  $\Omega_l^\prime(u_l,u_l^\prime)=\bigl(u_l^\prime,\Omega_l(u_l)\bigr)$
  \cite{Recknagel:2002qq}. This choice of automorphisms -- in particular the
  exchange of the group factors -- reflects the fact that the target space
  $G^{\text{new}}$ is defined as an {\em asymmetric} coset. Other choices of
  automorphisms would not lead to a geometry which is consistent with the
  identification \eqref{eq:Identification}. The coupling of closed strings to
  previously described branes shows that they are localized along the
  twisted conjugacy classes
  $\mc{C}_{f_l^\prime}^{U_l\times U_l}(\Omega_l^\prime)=\bigl\{\bigl(s_lf_l^\prime t_l^{-1},t_lf_l^\prime\Omega_l(s_l^{-1})\bigr)\bigr\}$
  with $f_l^\prime$ determined by $\rho_l$ as like before \cite{Felder:1999ka}.
  By taking the product of the two entries, this set
  projects down to a twisted conjugacy class $\mc{C}_{f_l}^{U_l}(\Omega_l)$
  in the group $U_l$ where $f_l=f_l^{\prime2}$.
\medskip
  
  Let us summarize these results. We have argued that the D-branes which are
  described by the boundary state \eqref{eq:BoundaryState} are localized
  along a direct product of twisted conjugacy classes in the space
  $U_n\times U_{n-1}\times U_{n-1}\times\cdots\times U_1\times U_1$. Due to
  their symmetry properties they descend to the set $G^{\text{new}}$ which
  provides a valid target space reinterpretation. As all elements of
  $G^{\text{new}}$ may be brought to the form \eqref{eq:Representative}, we
  just recover the expression \eqref{eq:TwistedConjugacy} for the geometry of
  twisted D-branes which entered the Lagrangian approach.

% -----------------------------------------------------------------------
% -----------------------------------------------------------------------
\subsection{\label{sc:Spectrum}The spectrum of open strings}
  
  Our next task is to determine the spectrum of open strings which can end on
  a D-brane which is described by the boundary state \eqref{eq:BoundaryState}.
  The result has to be compared with predictions which arise from geometric
  description afterwards. The calculation of closed string propagation between
  two boundary states of the form \eqref{eq:BoundaryState} by world sheet
  duality yields the open string Hilbert space
\begin{equation}
  \label{eq:BoundaryPartition}
  \mc{H}_{\rho\rho}
  \ =\ \bigoplus\:\bigl(n_{\nu_n}^{U_n}\bigr)_{\rho_n}^{\rho_n}
  \:\Biggl[\prod_{l=1}^{n-1}N_{\nu_l\sigma_l^+}^{\lambda_l}\bigl(n_{\lambda_l}^{U_l}\bigr)_{\rho_l}^{\rho_l}\Biggr]\:
  \mc{H}_{(\nu_n,\sigma_{n-1})}^{U_n/U_{n-1}}\otimes\cdots\otimes\mc{H}_{(\nu_2,\sigma_1)}^{U_2/U_1}\otimes\mc{H}_{\nu_1}^{U_1}\ \ ,
\end{equation}
  where we used the abbreviation $\rho=(\rho_1,\cdots,\rho_n)$ to denote the
  boundary label. The NIM-reps $n^{U_l}$ have been defined in eq.\ %
  \eqref{eq:NIMrep} and the numbers $N$ are
  fusion coefficients of the affine Lie algebras $\mf{\hat{u}}_{k_l}$. The
  calculation proceeds in the same way as those in \cite{Quella:2002ct}.
\medskip
  
  In the geometric picture only the ground state structure of the
  Hilbert space \eqref{eq:BoundaryPartition} can be recovered. Let us denote
  by $\mc{H}_{\rho\rho}^{(0)}$ the set of all ground states which are present
  in eq.\ \eqref{eq:BoundaryPartition}. Our aim is to find an explicit
  expression for this space which solely contains
  geometric information in the limit $k\to\infty$. The
  ground states of affine representations transform in a representation
  of the underlying simple Lie algebra which is usually denoted by the same
  symbol. In contrast, it is more difficult to give a geometrical meaning
  to the coset representations. All we can do is to determine the number of
  ground states they contain. The latter is given by the branching
  coefficients which describe the embedding of the associated horizontal
  subalgebras. These considerations provide a dictionary of how to extract
  geometrical information out of eq.\ \eqref{eq:BoundaryPartition}. We simply
  have to replace affine representations $\mc{H}_{\nu_1}^{U_1}$ by
  representation spaces $V_{\nu_1}^{U_1}$ of $\mf{u}_1$ and branching spaces
  $\mc{H}_{(\nu_l,\sigma_{l-1})}^{U_l/U_{l-1}}$ by branching coefficients
  of the embedding $\mf{u}_{l-1}\embin\mf{u}_l$. When represented as a $H$
  module we end up with the following expression,
\begin{equation}
  \label{eq:GroundStates}
  \mc{H}_{\rho\rho}^{(0)}
  \ =\ \bigoplus\:\bigl(n_{\nu_n}^{U_n}\bigr)_{\rho_n}^{\rho_n}
  \:\Biggl[\prod_{l=1}^{n-1}N_{\nu_l\sigma_l^+}^{\lambda_l}\bigl(n_{\lambda_l}^{U_l}\bigr)_{\rho_l}^{\rho_l}\ {b_{\nu_{l+1}}}^{\sigma_l}\Biggr]\:
  V_{\nu_1}^{U_1}\ \ ,
\end{equation}
  for the space of ground states. The geometric limit of
  a WZNW theory is obtained by sending the level $k$ of the affine Kac-Moody
  algebra $\mf{\hat{g}}_k$ to infinity. This automatically forces the
  levels of the Kac-Moody subalgebras $(\mf{\hat{u}}_l)_{k_l}$ also to tend to
  infinity. In this limit the fusion coefficients
  $N$ entering eq.\ \eqref{eq:BoundaryPartition}
  reduce to tensor product coefficients and also the NIM-reps
  $n^{U_l}$ have a natural geometrical meaning
  \cite{Alekseev:2002rj}.
\medskip
  
  Let us now turn our attention to the evaluation of D-brane spectra in the
  geometric picture.
  As was shown in \cite{Alekseev:1999bs,Alekseev:2000fd} for ordinary conjugacy
  classes and then generalized to the twisted case in \cite{Alekseev:2002rj}
  there is always a noncommutative geometry associated to these objects.
  In some sense this reflects the geometric limit of the noncommutative algebra
  of open string vertex operators. For the ground states this algebra indeed
  becomes coordinate independent in the large volume limit $k\to\infty$
  as their conformal dimensions tend to zero. For an arbitrary D-brane wrapped
  around the twisted conjugacy class
  $\mc{C}_{f_l}^{U_l}\bigl(\Omega_l\bigr)$ this noncommutative algebra
  admits an action of $U_l$ under which it decomposes into modules
  $V_{\nu_l}^{U_l}$ according to
\begin{equation}
  \label{eq:SingleAlgebra}
  \mc{A}\Bigl(\mc{C}_{f_l}^{U_l}\bigl(\Omega_l\bigr)\Bigr)
  \ =\ \bigoplus\ \bigl(n_{\nu_l}\bigr)_{\rho_l}^{\rho_l}\ V_{\nu_l}^{U_l}\ \ .
\end{equation}
  The same expression would be obtained for
  $\text{Inv}_{U_l}\mc{A}\bigl(\mc{C}_{f_l^\prime}^{U_l\times U_l}\bigl(\Omega_l^\prime\bigr)\bigr)$
  where the invariance is defined with respect to the identification 
  $(u_l^\prime,u_l)\sim(u_l^\prime\cdot t_l^{-1},t_l\cdot u_l)$ for
  $t_l\in U_l$ (cf.\ eq.\ \eqref{eq:Identification}).
  For our further discussion it is not necessary to know the detailed form
  of the matrices $n^{U_l}$. It is only important to note that the numbers
  $n^{U_l}$ entering eq.\ \eqref{eq:GroundStates} coincide with those in
  eq.\ \eqref{eq:SingleAlgebra} in the limit $k\to\infty$
  \cite{Alekseev:2002rj}.
\medskip
  
  The module structure of the algebra for a (direct) product of twisted
  conjugacy classes is given by the tensor product of the individual modules.
  We thus obtain the spectrum of open string ground states
\begin{equation}
  \label{eq:GeometricSpectrum}
  \mc{A}\Bigl(\mc{D}\bigl\{U_l,\Omega_l,f_l\bigr\}\Bigr)
  \ =\ \bigoplus_{\{\nu_l\}}\ \biggl[\prod_{l=1}^n\bigl(n_{\nu_l}\bigr)_{\rho_l}^{\rho_l}\biggr]\ V_{\nu_n}^{U_n}\otimes\cdots\otimes V_{\nu_1}^{U_1}\ \ ,
\end{equation}
  interpreted as a module of the group $U_n\times\cdots\times U_1$. In our
  approach of section \ref{sc:Lagrange}, the twisted conjugacy classes are not
  viewed as modules of the groups $U_l$ but as modules of the diagonally
  embedded $H=U_1$. This means that we should fully decompose the module
  \eqref{eq:GeometricSpectrum} with respect to $H$ in order to
  read off the spectrum of ground states belonging to the D-brane
  described by $\mc{D}\bigl\{U_l,\Omega_l,f_l\bigr\}$. In this way we get
  a number of additional branching and tensor product coefficients.
\medskip
  
  It is now straightforward to check the equality of the expressions on the
  right hand side of eqs.\ \eqref{eq:GroundStates} and
  \eqref{eq:GeometricSpectrum} in the limit $k\to\infty$, both considered as
  $H$ modules. In other words we have just proven the relation
  $\mc{A}\bigl(\mc{D}\bigl\{U_l,\Omega_l,f_l\bigr\}\bigr)\cong\mc{H}_{\rho\rho}^{(0)}$
  which expresses the agreement of the open string spectra obtained both
  from an algebraic and a geometric point of view, respectively. Let us
  emphasize that the geometric open string algebra
  $\mc{A}\bigl(\mc{D}\bigl\{U_l,\Omega_l,f_l\bigr\}\bigr)$ may not be
  identified with the algebra of functions on the D-brane world volume
  $\mc{D}\bigl\{U_l,\Omega_l,f_l\bigr\}\subset G$. Again this may be
  interpreted as an effect in favour of working with the new target space
  $G^{\text{new}}$. When
  considered as an object in $G$, the points in the D-brane world volume can be
  covered more than once. Obviously it is not possible to describe the new
  degrees of freedom which are associated to such multiple wrappings and
  superpositions of D-branes by the usual algebra of functions on the world
  volume $\mc{D}\bigl\{U_l,\Omega_l,f_l\bigr\}\subset G$. Similar observations
  have been discussed in \cite{Maldacena:2001ky}.

% -----------------------------------------------------------------------
% -----------------------------------------------------------------------
% -----------------------------------------------------------------------
\section{\label{sc:Applications}Applications}

  In this section we will apply our general results to advance the
  classification of D-branes in the target space $G=SL(2,\Real)\times SU(2)$.
  When lifted to the covering space $AdS_3$ of $SL(2,\Real)$ these
  provide us with non-factorizing and symmetry breaking D-branes in the
  string backgrounds $AdS_3\times S^3\times T^4$ and
  $AdS_3\times S^3\times S^3\times S^1$. The CFT description of WZNW models
  for noncompact groups is very intricate as their spectrum is continuous.
  In particular their analysis is not covered by \cite{Quella:2002ct}. We have
  therefore nothing new to say about the CFT side of constructing D-branes in
  these backgrounds. Nevertheless we conjecture that at least qualitatively our
  geometric analysis reflects the correct picture. Parts of our observations
  for factorizing D-branes in $SL(2,\Real)\times SU(2)$ are immediate
  consequences
  of earlier results which have been obtained by different methods
  \cite{Stanciu:1999nx,Figueroa-O'Farrill:2000ei,
  Bachas:2000fr,Maldacena:2001ky,Rajaraman:2001ew,Sarkissian:2002ie}.
  For the sake of completeness and in order to illustrate the power of our new
  geometric description we also review these cases. In particular we will show
  that our approach confirms the T-duality prediction for the geometry of
  symmetry breaking D-branes in $SL(2,\Real)$ and $SU(2)$
  \cite{Maldacena:2001ky,Sarkissian:2002ie}.
\medskip

  According to the general scheme we have first to find out all inequivalent
  embeddings of continuous subgroups into each of the constituents
  $SL(2,\Real)$ and $SU(2)$. In addition we have to classify all the
  automorphisms of these subgroups. Common subgroups of $SL(2,\Real)$ and
  $SU(2)$ can then be used to construct non-factorizing symmetry breaking
  D-branes.

% -----------------------------------------------------------------------
% -----------------------------------------------------------------------
\subsection{Preliminaries}

  The continuous subgroups of the product $SL(2,\Real)\times SU(2)$
  are easily classified. There are essentially two choices of embedding
  chains \eqref{eq:EmbeddingChain} which may be used for our construction.
  The first one is given by the maximal embedding
\begin{equation}
  \label{eq:EmbOne}
  H_1\ \times\ H_2\ \embin\ SL(2,\Real)\times SU(2)\ \ ,
\end{equation}
  where $H_1$ and $H_2$ equal one of the groups $U(1)$ or $\Real$.
  Without loss of generality we assume that $H_1$ is embedded into
  $SL(2,\Real)$ and $H_2$ is embedded into $SU(2)$. In this case the
  embedding map is given by $\epsilon^{H_1,SL}\times\epsilon^{H_2,SU}$.
  To save space we used the abbreviations $SL$ and $SU$ for $SL(2,\Real)$ and
  $SU(2)$, respectively. Before we dive into the discussion of these maps, let
  us first write down the second choice for a chain of embeddings.
  It is specified by
\begin{equation}
  \label{eq:EmbTwo}
  H\ \embin\ H\ \times\ H\ \embin\ SL(2,\Real)\times SU(2)\ \ .
\end{equation}
  Again, the symbol $H$ denotes a group of type $U(1)$ or $\Real$.
  The embedding map can be written as
  $\bigl[\epsilon^{H,SL}\times\epsilon^{H,SU}\bigr]\circ\epsilon^{H,H\times H}$.
  The decomposition \eqref{eq:AlgDecompNew} of chiral algebras which arises
  from \eqref{eq:EmbTwo} contains cosets of the form $\mc{A}(H/H)$
  if $H$ is completely embedded in one of the factors $H\times H$. As these are
  nasty to deal with on the algebraic level, we assume all possible embeddings
  in relation \eqref{eq:EmbTwo} to have the diagonal form
  $\epsilon^{H,H\times H}(h)=(h,h)$. The remaining freedom is then
  contained in the embedding maps $\epsilon^{H,SL}\times\epsilon^{H,SU}$
  from $H\times H$ to $SL(2,\Real)\times SU(2)$.
\medskip

  Our next task is to classify all embeddings and automorphisms of
  groups and subgroups belonging to the relations
  \eqref{eq:EmbOne} and \eqref{eq:EmbTwo}. Readers familiar with these
  issues may directly jump to section \ref{sc:ApplicationsSLSU}.
  The embeddings for the subgroup $SO(2)\cong U(1)$ are given by
\begin{equation*}
  \epsilon_n^{U,SL}\bigl(e^{i\phi}\bigr)
  \ =\ \epsilon_n^{U,SU}\bigl(e^{i\phi}\bigr)
  \ =\ \mat\cos n\phi&\sin n\phi\\-\sin n\phi&\cos n\phi\tam
  \qquad\quad\text{for } n\in\Integer\backslash\{0\}\ \ .
\end{equation*}
  In addition both groups admit an embedding of $\Real$. For
  $SL(2,\Real)$ we have two inequivalent choices one of which utilizes
  the covering of $U(1)$ by $\Real$. The latter also allows an embedding
  of $\Real$ into $SU(2)$. The explicit maps read
\begin{equation}
  \label{eq:EmbeddingsR}
  \epsilon_\alpha^{\Real,SL}(\lambda)
  \ =\ \mat e^{\alpha\lambda}&0\\0&e^{-\alpha\lambda}\tam\quad\text{and}\quad
  \tilde{\epsilon}_\beta^{\Real,SL}(\lambda)
  \ =\ \tilde{\epsilon}_\beta^{\Real,SU}(\lambda)\ =\ \mat\cos\beta\lambda&\sin\beta\lambda\\-\sin\beta\lambda&\cos\beta\lambda\tam\ \ .
\end{equation}
  Obviously, we have to demand that $\alpha$ and $\beta$ are non-vanishing.
\medskip

  The automorphisms are also easily classified. The group $SU(2)$ only admits
  inner automorphisms. According to the remarks at the end of section
  \ref{sc:Lagrange}, we may assume them to act trivially without any
  restrictions. In contrast, there exist outer automorphisms of the group
  $SL(2,\Real)$ which are based on the representative
\begin{equation}
  \label{eq:AutSL}
  \Omega_0^{SL}\mat a&b\\c&d\tam\ =\ \mat d&c\\b&a\tam\ =\ \mat0&1\\1&0\tam\mat a&b\\c&d\tam\mat0&1\\1&0\tam\ \ .
\end{equation}
  Note that this automorphism allows a representation as conjugation
  with the matrix
  $M=\left(\begin{smallmatrix}0&1\\1&0\end{smallmatrix}\right)$
  which is not an element of $SL(2,\Real)$.
  For the subgroups $U(1)$ and $\Real$ we obtain
\begin{align}
  \Omega_\pm^{U}\bigl(e^{i\phi}\bigr)&\ =\ e^{\pm i\phi}&&\text{and}&
  \Omega_\pm^{\Real}(\lambda)&\ =\ \pm\lambda\ \ .
\end{align}
  In fact, we could choose for $\Real$ the multiplication with an
  arbitrary non-zero real number. In our description we implicitly demanded,
  however, that the automorphisms should be consistent with the scalar
  product in the Lie algebra, i.e.\ they should not change the embedding
  index.\footnote{This is automatically satisfied for all semi-simple groups
  where outer automorphisms arise from Dynkin diagram symmetries.} This
  condition restricts our possibilities to the multiplication by $\pm1$.
  Finally, we have to discuss the automorphisms of $H\times H$. In this paper
  we restrict ourselves to automorphisms of the form
  $\Omega^{H\times H}=\Omega\circ\bigl[\Omega_1^H\times\Omega_2^H\bigr]$,
  where $\Omega:H\times H\to H\times H$ denotes a possible
  exchange of the two group factors and $\Omega_1^H,\Omega_2^H$ are two
  arbitrary automorphisms of $H$. If $H$ would have been semi-simple our
  choice of $\Omega^{H\times H}$ would have exhausted all possibilities.
  For our abelian groups $U(1)$ and $\Real$, there are more general choices
  but they are generically plagued with some unwanted features.\footnote{The
  scalar product in the corresponding Lie algebra would not be invariant.}%
\medskip

  We are now almost prepared to address the question of D-brane geometry
  in $SL(2,\Real)\times SU(2)$. All we still need is a better understanding
  of certain twisted conjugacy classes. Let $H$ be one of the subgroups $U(1)$
  or $\Real$. As both groups are abelian one immediatley obtains
\begin{equation*}
  \mc{C}_f^H\bigl(\Omega_\eta^H\bigr)
  \ =\ \left\{\begin{array}{cl}
         \{f\}&,\ \eta=+\\[1mm]
         H&,\ \eta=-\ \ .
       \end{array}\right.
\end{equation*}
  The element $f$ has to satisfy the symmetry property $\Omega_\eta^H(f)=f$.
  It is slightly more complicated to find an expression
  for the twisted conjugacy classes
  $\mc{C}^{H\times H}\bigl(\Omega^{H\times H}\bigr)$.
  Considering first the case with $\Omega=\id$ we easily obtain
\begin{equation}
  \label{eq:AutOne}
  \mc{C}_{(f_1,f_2)}^{H\times H}\bigl(\Omega^{H\times H}\bigr)\Bigr|_{\Omega=\id}
  \ =\ \left\{\begin{array}{cl}
         \{f_1\}\times\{f_2\}
           &,\ \Omega_1^H=\id,\ \Omega_2^H=\id\\[1mm]
         \{f_1\}\times H&,\ \Omega_1^H=\id,\ \Omega_2^H\neq\id\\[1mm]
         H\times\{f_2\}&,\ \Omega_1^H\neq\id,\ \Omega_2^H=\id\\[1mm]
         H\times H&,\ \Omega_1^H\neq\id,\ \Omega_2^H\neq\id\ \ .
       \end{array}\right. 
\end{equation}
  In this case the element $(f_1,f_2)$ has to satisfy $\Omega_i^H(f_i)=f_i$.
  With non-trivial twist, i.e.\ with $\Omega\neq\id$, we have the restrictions
  $f_1=\Omega_2^H(f_2)$ and $f_2=\Omega_1^H(f_1)$. It thus suffices to work
  with one label $f=f_1$ which satisfies $\Omega_2^H\circ\Omega_1^H(f)=f$.
  A straightforward analysis yields
\begin{equation}
  \label{eq:AutTwo}
  \mc{C}_f^{H\times H}\bigl(\Omega^{H\times H}\bigr)\Bigr|_{\Omega\neq\id}
  \ =\ \left\{\begin{array}{cl}
         \qquad H\times H&,\ \Omega_1^H\neq\bigl(\Omega_2^H\bigr)^{-1}\text{ and }\Omega\neq\id\\[1mm]
         \bigl\{\bigl(sf,\Omega_1^H(fs^{-1})\bigr)\bigr|s\in H\bigr\}
           &,\ \Omega_1^H=\bigl(\Omega_2^H\bigr)^{-1}\text{ and }\Omega\neq\id\ \ .
       \end{array}\right. 
\end{equation}
  This concludes our presentation of the necessary tools for the determination
  of symmetry breaking D-branes in $SL(2,\Real)\times SU(2)$.

% -----------------------------------------------------------------------
% -----------------------------------------------------------------------
\subsection{\label{sc:ApplicationsSLSU}Symmetry breaking D-branes in $SL(2,\Real)$ and $SU(2)$ -- a review}

  This subsection will be used to introduce some notation which is necessary
  to describe the geometry of $SL(2,\Real)$ and $SU(2)$. In addition we will
  find that our approach is in agreement with recent  results for symmetry
  breaking D-branes in these groups which have been obtained using
  T-duality \cite{Maldacena:2001ky,Rajaraman:2001ew,Sarkissian:2002ie}.
  
% -----------------------------------------------------------------------
\subsubsection{\label{sc:ApplicationsSL}Symmetry breaking D-branes in $SL(2,\Real)$}

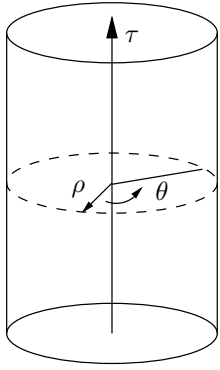
\begin{figure}
  \centerline{\input{SLCoord.pstex_t}}
  \caption{\label{fig:SL}Parametrization of $SL(2,\Real)$.}
\end{figure}

  The group $SL(2,\Real)$ may be described as a subspace of
  four dimensional flat space. In this parametrization the connection
  to the matrix form is given by
\begin{equation}
  \label{eq:GeoSLOne}
  \mat X_0+X_3&X_1+X_2\\X_1-X_2&X_0-X_3\tam
  \qquad\text{subject to}\qquad X_0^2-X_1^2+X_2^2-X_3^2\ =\ 1\ \ .
\end{equation}
  It is convenient to introduce cylindrical coordinates $r,\theta$
  and a periodic time $\tau$. These take values in the domains
  $r\in[0,\infty[$ and $\theta,\tau\in[0,2\pi[$. The precise relation
  to the previous parametrization is given by
\begin{equation}
  \label{eq:GeoSLTwo}
  X_0+iX_2\ =\ e^{i\tau}\cosh r
  \qquad\text{ and }\qquad
  X_3+iX_1\ =\ e^{i\theta}\sinh r\ \ .
\end{equation}
  In the cylindrical coordinates the manifold $SL(2,\Real)$ may be depicted
  as in figure \ref{fig:SL} with top and bottom of the cylinder identified.
  The covering space $AdS_3$ is obtained by resolving the periodicity of time,
  i.e. by extending its range to $\tau\in\Real$. The region $r\to\infty$
  describes the boundary of $AdS_3$.
\medskip

\begin{figure}
  \centerline{\input{SLTypC1.pstex_t}\qquad\input{SLTypC2.pstex_t}\qquad\input{SLTypC3.pstex_t}\qquad\input{SLTypT1.pstex_t}}
  \caption{\label{fig:SLMax}Representatives of maximally symmetric D-branes in
  $SL(2,\Real)$. From left to right we have the following types: point-like,
  $dS_2$, $H_2$ and $AdS_2$ branes.}
\end{figure}
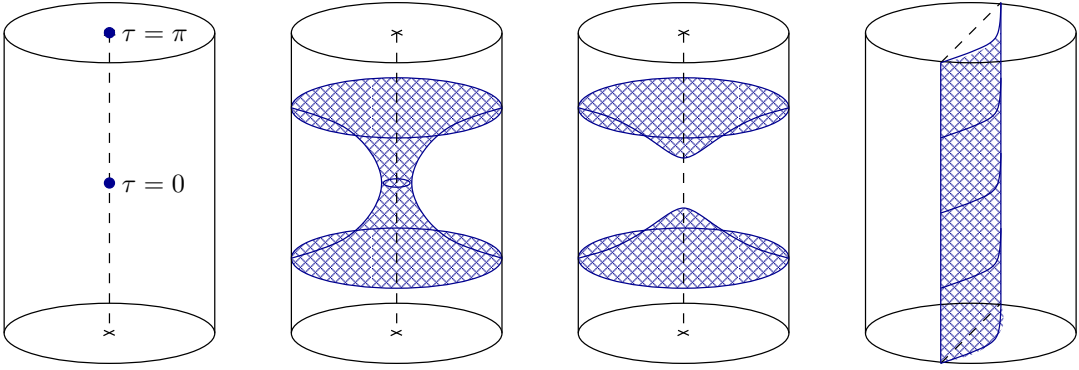

  Maximally symmetric D-branes belong to twisted conjugacy classes of
  $SL(2,\Real)$. For ordinary conjugacy classes all elements are mapped to the
  same number by taking the trace. Fixing $X_0$ to some value
  $C\in\Real$ while putting no constraints on the other coordinates therefore
  gives a first rough classification. The resulting submanifold may be
  disconnected showing that only demanding $X_0=C$ does not lead to a complete
  solution of the classification problem. It is nevertheless useful to work
  with this description. The equation $X_0=\cos\tau\cosh r=C$ admits
  very different types of solutions depending on whether $|C|<1$ or
  $|C|>1$. For $|C|>1$ one recovers $dS_2$ branes while for $|C|<1$
  branes are obtained which are localized on hyperbolic planes $H_2$.
  In the limit $|C|\to1$ they degenerate to two
  instantonic point-like D-branes at $\tau=0,\pi$ which are associated to the
  center of $SL(2,\Real)$ and others sitting on the light cone. Representatives
  of this zoo of
  conjugacy classes are visualized in figure \ref{fig:SLMax}. These results
  have already been found in \cite{Stanciu:1999nx,Bachas:2000fr}. It was argued
  in \cite{Bachas:2000fr} that all these D-branes are unphysical. The
  $H_2$ and the point-like branes are instantonic objects while the $dS_2$
  branes are spoiled by a supercritical electrical field.
\medskip

  Twisted conjugacy classes coming from the automorphism \eqref{eq:AutSL} are
  classified by the relation $\tr\bigl(Mg\bigr)=2X_2=2C$. According to eq.\ %
  \eqref{eq:GeoSLTwo} this translates into $C=\sin\theta\sinh r$. In this
  situation there is no need to distinguish different cases. All these twisted
  conjugacy classes describe $AdS_2$ branes which are invariant under time
  translations and extend to the boundary of $AdS_3$ at $\theta=0,\pi$
  \cite{Bachas:2000fr}. They are illustrated in the right-most picture of
  figure \ref{fig:SLMax}.
\medskip
  
  Let us now turn to the description of symmetry breaking D-branes. According
  to the expression \eqref{eq:TwistedConjugacy} we may multiply the twisted
  conjugacy classes of $SL(2,\Real)$ by a twisted conjugacy class
  $\mc{C}_f^H\bigl(\Omega^H\bigr)$ of $H=U(1)$ or $H=\Real$. For $\Omega^H=\id$
  the latter are point-like. This induces a shift of the original
  D-brane. The situation is more interesting if $\Omega^H\neq\id$. In this case
  the twisted conjugacy class reduces to $H$ itself and one has to consider the
  superposition of all shifted images.
\medskip
  
  This analysis is particularly simple for $H=U(1)$. In this case the
  multiplication of an element $(r,\theta,\tau)$ of $SL(2,\Real)$ with
  an element $e^{i\lambda}$ of $U(1)$ just induces the simultaneous rotation
  $(\theta,\tau)\mapsto(\theta\pm n\lambda,\tau\pm n\lambda)$ of angle and time
  coordinate. The sign and the wrapping number $n$ are fixed by the choice of
  twist $\Omega^{SL}$ and embedding $\epsilon^{U,SL}$.
  As the twisted conjugacy class is given by the whole $U(1)$, one
  may immediately evaluate the geometry of the resulting D-branes. By rotation
  of the $dS_2$ and the $AdS_2$ branes one obtains D-branes which fill all
  space outside a cylinder of radius $r_0=\text{arcosh}|C|$ or
  $r_0=\text{arsinh}|C|$, respectively. For degenerate cases
  they provide space-filling branes similar to those arising
  from the rotation of $H_2$ branes. If one rotates the $0$-branes on the
  other hand these sweep out the axis $r=0$. To get some impression of these
  geometries we visualized all four of them in figure \ref{fig:SLMax}.
  Let us emphasize that the generic point in the world volume of rotated
  $dS_2$ and $AdS_2$ branes is covered twice.
\medskip
  
  For $H=\Real$ we have to distinguish two embeddings \eqref{eq:EmbeddingsR}.
  The usage of $\tilde{\epsilon}_\beta^{\Real,SL}$ gives essentially the same
  result as for $U(1)$. For $\epsilon_\beta^{\Real,SL}$, in contrast, the
  discussion becomes quite involved as the shift acts in a very intricate way
  --  at least in our coordinates $(r,\theta,\tau)$.
  To get an idea of what is going, on let us consider the case where
  the conjugacy class of $SL(2,\Real)$ reduces to a point. The D-brane
  is then parametrized by matrices of the form
  $\text{diag}(\pm e^\lambda,\pm e^{-\lambda})$ with $\lambda\in\Real$.
  It turns out that these are instantonic D1-branes localized at times
  $\tau=0,\pi$, respectively, and running all the way from $r=0$ to
  $r=\infty$ in the directions $\theta=0,\pi$. They do not seem to make
  sense physically and we will not discuss them
  in more detail. Notice that most of the symmetry breaking D-branes which
  have been described in this section already appeared in
  \cite{Maldacena:2001ky,Rajaraman:2001ew,Sarkissian:2002ie}.
  
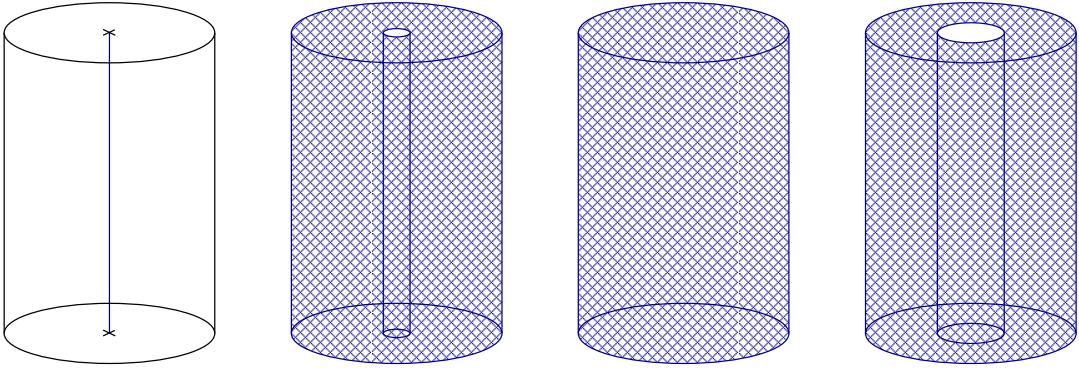
\begin{figure}
  \centerline{\input{SLTypS1.pstex_t}\qquad\input{SLTypS2.pstex_t}\qquad\input{SLTypS3.pstex_t}\qquad\input{SLTypS4.pstex_t}}
  \caption{\label{fig:SLBreaking}Certain classes of symmetry breaking D-branes
  on $SL(2,\Real)$. They are obtained from those in figure \ref{fig:SLMax} by a
  simultaneous $(\theta,\tau)$-rotation.}
\end{figure}

% -----------------------------------------------------------------------
\subsubsection{\label{sc:ApplicationsSU}Symmetry breaking D-branes in $SU(2)$}

\begin{figure}
  \centerline{\input{SUTypC1.pstex_t}\qquad\input{SURot.pstex_t}\qquad\input{SUTypS1.pstex_t}}
  \caption{\label{fig:SUMax}Maximally symmetric and symmetry breaking D-branes
  on $SU(2)$. The latter arise from the rotation indicated in the
  central picture. They generically cover a $3$-dimensional subset of $S^3$
  but leave open a ``window'' of a certain size.}
\end{figure}
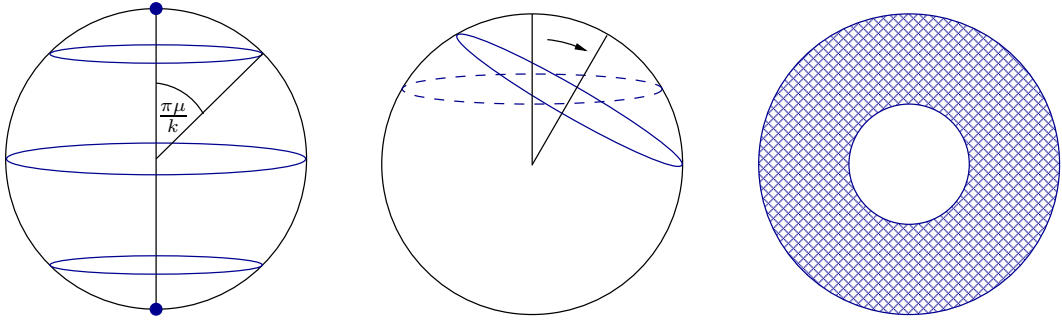

  The group manifold $SU(2)$ may be realized as a subset of $\Complex^2$.
  In this parametrization the elements are described by a matrix
  $\left(\begin{smallmatrix} z_1&z_2\\-\bar{z}_2&\bar{z}_1\end{smallmatrix}
   \right)$
  subject to the condition $|z_1|^2+|z_2|^2=1$.
  Maximally symmetric D-branes are localized along quantized conjugacy classes
  \cite{Alekseev:1998mc,Felder:1999ka}.
  For a WZNW theory at level $k$ we have $k+1$ spheres $S^2$ which
  sit at the special values $\text{Re}(z_1)=\cos\frac{\pi\mu}{k}$ with
  $\mu=0,\cdots,k$. For $\mu=0,k$ they degenerate to points. An illustration
  of these facts is given on the left hand side of figure \ref{fig:SUMax}.
\medskip
  
  To describe symmetry breaking D-branes we have to multiply these conjugacy
  classes by twisted conjugacy classes of $U(1)$ or $\Real$. Choosing a
  trivial automorphism amounts to a shift like before. When considering a
  non-trivial automorphism we have to take the union of all these shifted
  images. We will not write down the explicit expressions but only refer to
  the illustration on the right hand side of figure
  \ref{fig:SUMax}. The symmetry breaking D-branes are either $1$- or
  $3$-dimensional. While the first ones are circular, the latter cover most of
  the group but generically leave some parts uncovered. Let us emphasize that
  we also find a  space-filling brane by considering the conjugacy classes of
  $SU(2)$ with $\mu=k/2$ for even values of $k$. It is remarkable to note that
  a generic point of all these D-branes is covered twice. This observation is
  related to the fact that the space-filling brane can be further resolved
  into two single branes. These results are not new but have already been
  obtained in \cite{Maldacena:2001ky,Sarkissian:2002ie} with different methods.
  
% -----------------------------------------------------------------------
% -----------------------------------------------------------------------
\subsection{On the hierarchy of D-branes in $SL(2,\Real)\times SU(2)$}

  When constructing D-branes in the product geometry $SL(2,\Real)\times SU(2)$,
  it is convenient to distinguish three cases which belong to qualitatively
  different classes of automorphisms for the subgroups appearing in the two
  embedding chains \eqref{eq:EmbOne} and \eqref{eq:EmbTwo}. The discussion
  of this classification will be the subject of the following three
  subsections.

% -----------------------------------------------------------------------
\subsubsection{\label{sc:FactorizingDBranes}Factorizing D-branes}

  Factorizing D-branes arise from the embedding chain \eqref{eq:EmbOne}.
  For $H=H_1=H_2$ we further have to assume that the automorphism of
  $H\times H$ involved in the construction of the D-branes does not contain the
  exchange of
  the group factors. Under these circumstances the decomposition
  \eqref{eq:AlgDecompNew} reduces to a separate decomposition of
  $\mc{A}\bigl(SL(2,\Real)\bigr)$ and $\mc{A}\bigl(SU(2)\bigr)$ in both
  holomorphic and antiholomorphic degrees of freedom.
\medskip

  Let us discuss the geometry of these D-branes now. According to the general
  expression \eqref{eq:TwistedConjugacy} they are localized along the product
\begin{equation*}
  \Bigl[\mc{C}_{f_1}^{SL}\bigl(\Omega^{SL}\bigr)\cdot\Omega^{SL}\circ\epsilon^{H_1,SL}\Bigl(\mc{C}_{f_2}^{H_1}\bigl(\Omega^{H_1}\bigr)\Bigr)\Bigr]
  \ \times\ %
  \Bigl[\mc{C}_{f_3}^{SU}\bigl(\id\bigr)\cdot\epsilon^{H_2,SU}\Bigl(\mc{C}_{f_4}^{H_2}\bigl(\Omega^{H_2}\bigr)\Bigr)\Bigr]\ \ .
\end{equation*}
  This factorized geometry is completely under control using the dictionary
  which has been provided in section \ref{sc:ApplicationsSLSU}.
  The dimensions of these D-branes range from
  $0$ to $6$, the shape from point-like to
  space-filling. We will not bother to discuss these D-branes any further but
  focus our attention on the description of non-factorizing D-branes.

% -----------------------------------------------------------------------
\subsubsection{Non-factorizing D-branes from diagonal embedding}

  The first type of non-factorizing D-branes is obtained by using the
  embedding chain \eqref{eq:EmbTwo} and choosing an automorphism
  $\Omega^{H\times H}$ which does not involve an exchange of the two factors.
  In other words we demand $\Omega=\id$. The geometry associated to this kind
  of symmetry breaking D-brane is described by the product
\begin{equation}
  \label{eq:aux7}
  \Bigl[\mc{C}_{f_1}^{SL}\bigl(\Omega^{SL}\bigr)\times\mc{C}_{f_2}^{SU}\bigl(\id\bigr)\Bigr]\ \cdot\ \epsilon_1\Bigl(\mc{C}_{(f_3,f_4)}^{H\times H}\bigl(\Omega^{H\times H}\bigr)\Bigr)\ \cdot\ \epsilon_2\Bigl(\mc{C}_{f_5}^H\bigl(\Omega^H\bigr)\Bigr)\ \ ,
\end{equation}
  where the embeddings have been abbreviated by
  $\epsilon_1=\bigl(\Omega^{SL}\circ\epsilon^{H,SL}\bigr)\times\epsilon^{H,SU}$ and
  $\epsilon_2=\bigl[\Omega^{SL}\circ\epsilon^{H,SL}\times\epsilon^{H,SU}\bigr]\circ\Omega^{H\times H}\circ\epsilon^{H,H\times H}$. Let us start our discussion with a
  given product of conjugacy classes of
  $SL(2,\Real)$ and $SU(2)$. As can be seen from eq.\ \eqref{eq:AutOne}
  the effect of the multiplication with
  $\mc{C}_{(f_3,f_4)}^{H\times H}\bigl(\Omega^{H\times H}\bigr)$ is a
  combination of a factorized smearing as described in section
  \ref{sc:ApplicationsSLSU}
  and a translation. The effect of the multiplication
  with $\mc{C}_{f_5}^H\bigl(\Omega^H\bigr)$ is more interesting as it can
  provide the reason for non-factorizability. If this conjugacy class is
  $0$-dimensional it shifts the whole D-brane by a constant amount leaving
  factorizability unaffected. On the other hand it may reduce to $H$ itself.
  Under these circumstances one obtains
  a continuous superposition of shifted D-branes. Due to the diagonal embedding
  of $H$ into $H\times H$, the shift acts on both factors $SL(2,\Real)$
  and $SU(2)$ {\em simultaneously}. This feature is responsible for
  non-factorizability.
\medskip

  The discussion of the geometry of these D-branes becomes rather
  involved in the general case. We prefer to illustrate our considerations
  in two simple examples. Assume first that $H=\Real$,
  $\epsilon^{H,SL}=\tilde{\epsilon}_\alpha^{\Real,SL}$ and that we set
  $f_1=f_2=f_3=f_4=e$, $\Omega^{SL}=\id$ and $\Omega_1^H=\Omega_2^H=\id$.
  This implies that the twisted conjugacy classes of $SL(2,\Real)$, $SU(2)$
  and $H\times H$ entering \eqref{eq:aux7} reduce to unit elements. In order
  to obtain a non-trivial result we choose $\Omega^H\neq\id$ such that the
  remaining twisted conjugacy class in eq.\ \eqref{eq:aux7} is given by $H$.
  After its embedding into $SL(2,\Real)\times SU(2)$ one recovers the curve
\begin{equation*}
  \Biggl(\mat\cos\alpha\lambda&\sin\alpha\lambda\\-\sin\alpha\lambda&\cos\alpha\lambda\tam\ ,\ \mat\cos\beta\lambda&\sin\beta\lambda\\-\sin\beta\lambda&\cos\beta\lambda\tam\Biggr)
  \qquad\text{ with }\lambda\in\Real\ \ .
\end{equation*}
  The numbers $\alpha,\beta$ specifiy the individual embeddings of $\Real$
  into the group factors. A closer comparison of this expression with
  eqs.\ (\ref{eq:GeoSLOne},\,\ref{eq:GeoSLTwo}) shows that $\alpha\lambda$
  may be identified with time $\tau$. This configuration thus describes
  a number of D-particles each having a circular trajectory in the group
  factor $SU(2)$ while sitting on the axis $r=0$ of $SL(2,\Real)$. The number
  of D-particles is determined by the relative values of $\alpha$ and $\beta$.
  If the ratio is irrational one obtains an infinite number of particles
  which form a dense set in $SU(2)$ at each instance of time. The appearance
  of multiple D-particles is due to the periodicity of time in $SL(2,\Real)$.
  This artefact disappears on the covering space $AdS_3$ which has a
  non-compact time-coordinate. Obviously, it is straightforward to generalize
  the previous idea to $2$-spheres which are rotating in the $SU(2)$ factor in
  the evolution of time.
\medskip

  In our second example we choose $\Omega^{SL}=\Omega_0^{SL}$, but still fix
  $H=\Real$, $\epsilon^{H,SL}=\tilde{\epsilon}_\alpha^{\Real,SL}$,
  $f_3=f_4=e$, $\Omega_1^H=\Omega_2^H=\id$ and $\Omega^H\neq\id$.
  The set \eqref{eq:aux7} is obtained from the product
  $AdS_2\times S^2$ by performing simultaneous shifts in both factors. If
  we focus only on the $AdS_2$-part for a moment we already know the resulting
  geometry from section \ref{sc:ApplicationsSL}. It is given by all points
  $(r,\theta,\tau)$ which satisfy $r\geq r_0=\text{arsinh}|C|$ for some
  constant $C$. These points are generically not located on the twisted
  conjugacy class of $SL(2,\Real)$ we have started from. We thus have to
  decompose them into an element $(r,\theta^\prime,\tau^\prime)$ and a shift
  $\lambda$ such that
  $(\theta,\tau)=(\theta^\prime+\lambda,\tau^\prime+\lambda)$
  and $\sin\theta^\prime\sinh r=C$, i.e.\ such that
  $(r,\theta^\prime,\tau^\prime)$ is an element of the twisted conjugacy class.
  With every solution $\theta_1^\prime$ we have another one
  $\theta_2^\prime=\pi-\theta_1^\prime$. In the exceptional case $r=r_0$
  we have only one solution $\theta_1^\prime=\theta_2^\prime=\pi/2$.
  For $r=r_0=0$ the angle can be chosen arbitrary. For simplicity we shall
  assume $r_0>0$ in what follows.
  
  The two shifts $\lambda_{1/2}$ associated to the angles
  $\theta_{1/2}^\prime$ have to come from the embedding
  $\tilde{\epsilon}_\alpha^{\Real,SL}$ of $\xi\in\Real$ in $SL(2,\Real)$. As
  $\alpha\xi_{1/2}$ is only defined modulo $2\pi$ there are several choices
  $\xi_{1/2}^{(l)}=(\lambda_{1/2}+2\pi\alpha l)/\alpha$ of elements in $\Real$
  which may be used to recover these shifts. These elements have to be used
  to implement the shift on the $SU(2)$-part. Using the embedding
  $\tilde{\epsilon}_\beta^{\Real,SU}$ as before, these shifts are determined
  by the angles $\beta(\lambda_{1/2}+2\pi\alpha l)/\alpha$. For
  $\alpha=\beta=1$ we arrive at the following picture. The D-brane
  in $SL(2,\Real)\times SU(2)$ is parametrized by points $(r,\theta,\tau)$
  in $SU(2,\Real)$ with $r\geq r_0$. Over each of these points one has two
  spheres $S^2$ which are generated out of the conjugacy class of $SU(2)$ by
  the action of the shifts $\lambda_{1/2}(r)$. In the limiting regimes
  $r\to r_0$ and $r\to\infty$ the two-spheres move closer and closer. For more
  general choices of $\alpha$ and $\beta$ the number of two-spheres over each
  point in $SL(2,\Real)$ may be larger.

% -----------------------------------------------------------------------
\subsubsection{Non-factorizing D-branes from group interchanging twists}

  One can also consider the embedding chain \eqref{eq:EmbTwo} having a
  twist $\Omega\neq\id$ of the $H\times H$ subgroup which interchanges
  the two factors. From the geometric point of view there are no difficulties
  in doing so. But at this point we have to remember that geometric
  automorphisms should be related to gluing automorphisms of currents on
  the level of chiral algebras. These chiral algebras
  are generated by affine Kac-Moody algebras and thus have the additional
  notion of a level. Consequently, the chiral algebras $\mc{A}(H)$
  may differ depending on whether they come from an embedding of $H$ in
  $SL(2,\Real)$ or $SU(2)$. The geometric twist of $H\times H$ can only be
  lifted to an automorphism of $\mc{A}_{SL}(H)\oplus\mc{A}_{SU}(H)$
  if these two algebras agree. This enforces some constraints on the relative
  size -- the levels -- of $SL(2,\Real)$ and $SU(2)$ and the embeddings
  one uses.
\medskip
  
  After these remarks we can proceed as in the previous section.
  The geometry which belongs to our present choice of embedding chain
  may be read off from the product
\begin{equation}
  \label{eq:aux8}
  \Bigl[\mc{C}_{f_1}^{SL}\bigl(\Omega^{SL}\bigr)\times\mc{C}_{f_2}^{SU}\bigl(\id\bigr)\Bigr]\ \cdot\ \epsilon_1\Bigl(\mc{C}_{f_3}^{H\times H}\bigl(\Omega^{H\times H}\bigr)\Bigr)\ \cdot\ \epsilon_2\Bigl(\mc{C}_{f_4}^H\bigl(\Omega^H\bigr)\Bigr)\ \ ,
\end{equation}
  where we used the abbreviations
  $\epsilon_1=\bigl(\Omega^{SL}\circ\epsilon^{H,SL}\bigr)\times\epsilon^{H,SU}$ and
  $\epsilon_2=\bigl[\Omega^{SL}\circ\epsilon^{H,SL}\times\epsilon^{H,SU}\bigr]\circ\Omega^{H\times H}\circ\epsilon^{H,H\times H}$.
  The discussion of the conjugacy class $\mc{C}_{f_4}^H\bigl(\Omega^H\bigr)$
  gives no new insights compared to the previous section. Despite of this fact
  there is still
  a significant qualitative difference, as we are now allowed to work with
  the expressions \eqref{eq:AutTwo} for the conjugacy classes
  $\mc{C}_{f_3}^{H\times H}\bigl(\Omega^{H\times H}\bigr)$. While the first
  possibility implies the usual factorized smearing, the second induces
  a superposition of simultaneous shifts in the two group factors similar to
  those arising possibly from $\mc{C}_{f_4}^H\bigl(\Omega^H\bigr)$.
  It is an interesting question to see whether the joint action of two
  independent simultaneous shifts will lead to new features.
\medskip

  To illustrate these considerations we choose a setup where $H=\Real$,
  $\epsilon^{H,SL}=\tilde{\epsilon}_\alpha^{\Real,SL}$,
  $f_1=f_2=e$, $\Omega^{SL}=\id$ and
  $\Omega_1^H=\Omega_2^H=\Omega_\eta^H$ for $\eta=\pm1$ as well as
  $\Omega^H\neq\id$. The product of the embedding of the two twisted conjugacy
  classes \eqref{eq:aux8} is parametrized by two real numbers
  $\lambda,\lambda^\prime$ and reads
\begin{equation*}
  \Biggl(\mat\cos\psi&\sin\psi\\-\sin\psi&\cos\psi\tam\ ,\ \mat\cos\psi^\prime&\sin\psi^\prime\\-\sin\psi^\prime&\cos\psi^\prime\tam\Biggr)
\end{equation*}
  with $\psi=\alpha(\eta\lambda+\lambda^\prime+f_3)$ and
  $\psi^\prime=\beta\eta(\lambda-\lambda^\prime+f_3)$.
  We recognize that the joint action of simultaneous shifts for $\eta=1$
  leads to a factorized structure again as both $\psi$ and $\psi^\prime$
  are independent. For $\eta=-1$ they become dependent and one recovers
  a shifted version of the already familiar non-factorizing D-brane
  instead.

% -----------------------------------------------------------------------
% -----------------------------------------------------------------------
% -----------------------------------------------------------------------
\section{Conclusions}

  In the present work we constructed the boundary WZNW functional for D-branes
  which are localized along products of generalized twisted conjugacy classes
  of subgroups $U_l$ of $G$ which are organized in an embedding chain
  of the form \eqref{eq:EmbeddingChain}. The action functional was shown to
  be invariant under the action of a continuous subgroup. The D-branes
  have been identified as geometric counterparts of symmetry breaking boundary
  states constructed algebraically in \cite{Quella:2002ct}. Our results yield
  a huge hierarchy of D-branes in group manifolds which are under complete
  control both from an algebraic and a geometric point of view.
\medskip
  
  We focused on the example of $SL(2,\Real)\times SU(2)$ to expedite the
  classification of D-branes in this background. Although the
  groups $SL(2,\Real)$ and $SU(2)$ only possess a small number of subgroups
  which in addition are abelian, we perceived a glimpse of the possibilities
  which open up for
  more complicated backgrounds. In particular we are now able to work out
  the geometry of D-branes in product groups which do not necessarily
  factorize \cite{Quella:2002ct} and which do not originate from a twist of
  two group factors with equal size \cite{Figueroa-O'Farrill:2000ei,
  Recknagel:2002qq}. The same ideas allow us to investigate defect lines in
  $1+1$-dimensional quantum field theories \cite{Quella:2002ct}.
  Such defects are for example induced on the holographic dual
  of $AdS_3$ by D-branes in its interior which extend to the boundary
  \cite{Karch:2000gx,DeWolfe:2001pq,Bachas:2001vj,Erdmenger:2002ex}.
  It would be interesting to see whether our observations for D-branes in the
  non-compact group $SL(2,\Real)\times SU(2)$ can be confirmed by an algebraic
  construction of boundary states.%
\medskip
  
  Our algebraic analysis implied the necessity of target space
  reinterpretation in order to resolve certain ambiguities in the geometric
  description of symmetry breaking D-branes. 
  In this way we revealed a striking connection between string theory
  on group manifolds and asymmetrically gauged coset spaces
  \cite{Guadagnini:1987ty,Bars:1992pt}. Although there has been some
  work recently on D-branes in asymmetrically
  gauged coset theories \cite{Walton:2002db}, this subject is still
  not well-developed. Nevertheless there exist several important examples where
  such backgrounds play a prominent role. Let us only mention at this place
  the Nappi-Witten background \cite{Nappi:1992kv,Elitzur:2002rt} and the
  $T^{pq}$ spaces \cite{Pando-Zayas:2000he}. The first of them provides an
  interesting example of a cosmological background with big-bang and
  big-crunch singularity while the second is related to the base of a conifold
  for $p=q=1$ \cite{Candelas:1990js,Klebanov:1998hh}.
\medskip

  The relation between string theory on group manifolds and asymmetrically
  gauged coset spaces may be extended beyond the simple observation of
  target space reinterpretation.
  Our results indeed admit an immediate generalization to symmetry breaking
  D-branes in asymmetric coset models by simply gauging a subgroup of the
  $H$ symmetry which was preserved by the symmetry breaking D-branes in the
  group manifold. This is completely obvious from the geometric and the
  Lagrangian point of view. Gauging a subgroup of $G$ makes necessary the
  introduction of a gauge field interaction term
  \cite{Gawedzki:1988hq,Karabali:1989au,Bars:1992pt}. The boundary
  contribution, however, remains unaffected by this extra term.
  A first step towards a general and comprehensive description of
  asymmetrically gauged coset models in terms of an algebraic and geometric
  analysis including D-branes will follow in \cite{TQVS:Unpublished}.
\medskip
  
  To summarize, our results along with those of \cite{Quella:2002ct}
  provide a powerful tool to push forward the programme of classifying
  D-branes in a given background. It would be very interesting
  to apply these methods to examples which possess a richer structure of
  non-abelian subgroups. Let us emphasize the general feature that the
  dimension of the D-branes increases with the amount of symmetry breaking we
  enforce. In particular we have argued that one should generically be able to
  recover space-filling D-branes.
  To obtain a complete picture of D-branes in group manifolds and coset models
  it remains to analyze the stability and the dynamics of symmetry
  breaking D-branes along the lines of \cite{Bachas:2000ik,Bordalo:2001ec} and
  \cite{Alekseev:2000fd,Fredenhagen:2001kw,Fredenhagen:2002qn,Alekseev:2002rj}.
  In this context one would also like to address the question of D-brane
  charges and K-theory \cite{Fredenhagen:2000ei,Maldacena:2001xj}. This
  would require an extension of our results to include supersymmetry.  

% -----------------------------------------------------------------------
\subsubsection*{Acknowledgements}

  The author is grateful to A.\ Yu.\ Alekseev, S.\ Fredenhagen, S.\ Ribault
  and V.\ Schomerus for useful discussions and comments on the manuscript.
  This paper was strongly influenced by the collaboration with V.\ %
  Schomerus on asymmetrically gauged coset theories \cite{TQVS:Unpublished}.
  The present work was financially supported by the Studienstiftung des
  deutschen Volkes.

\begin{appendix}
% -----------------------------------------------------------------------
% -----------------------------------------------------------------------
% -----------------------------------------------------------------------
\section{Computational details}

  In this appendix we will provide the computational details which have
  been omitted in section \ref{sc:Lagrange}. It is convenient to choose
  a more general framework and to generalize the notion of twisted conjugacy
  class. We will then be able to include recent proposals of
  \cite{Walton:2002db} in our description.
\medskip

  In the general approach we start with a family $U_l$ ($l=1,\ldots,n$) of
  continuous subgroups $H\embin U_l\embin G$ which do not necessarily satisfy
  the embedding chain property \eqref{eq:EmbeddingChain}.
  To each of these subgroups we associate three embeddings
  $\epsilon^{HU_l}:H\to U_l$ and $\epsilon_{L/R}^{U_lG}:U_l\to G$.
  The indices $L/R$ stand for left and right, respectively. We then
  define generalized twisted conjugacy classes of $U_l$ in $G$ by
\begin{equation*}
  \mc{C}_{\tilde{f}_l}^{U_l,G}\Bigl(\epsilon_{L}^{U_lG},\epsilon_{R}^{U_lG}\Bigr)
  \ =\ \Bigl\{\,c_l=\epsilon_{L}^{U_lG}\bigl(s_l\bigr)\cdot\tilde{f}_l\cdot\epsilon_{R}^{U_lG}\bigl(s_l^{-1}\bigr)\,\Bigr|\,s_l\in U_l\Bigr\}\ \ .
\end{equation*}
  These generalized twisted conjugacy classes admit an action of $H$ under
  which they transform as
\begin{equation}
  \label{eq:Trafo}
  s_l\ \mapsto\ \epsilon^{HU_l}(h)\cdot s_l
  \quad\quad\Rightarrow\quad\quad
  c_l\ \mapsto\ \epsilon_{L}^{U_lG}\circ\epsilon^{HU_l}\bigl(h\bigr)\cdot c_l\cdot\epsilon_{R}^{U_lG}\circ\epsilon^{HU_l}\bigl(h^{-1}\bigr)\ \ .
\end{equation}
  By putting $\epsilon_{L}^{U_lG}=\epsilon^{U_lG}$,
  $\epsilon_{R}^{U_lG}=\epsilon_\Omega^{U_lG}\circ\Omega_l$
  and $\tilde{f}_l=\epsilon^{U_lG}(f_l)$
  we could recover ordinary twisted conjugacy classes in this more general
  framework, i.e.\ the setup of section \ref{sc:Lagrange}.
  One easily verifies that the set
  $\mc{D}\bigl\{U_l,\epsilon_{L/R}^{U_lG},\tilde{f}_l\bigr\}$ which is
  generated by the following product of generalized twisted conjugacy classes,
\begin{equation}
  \label{eq:ConjugacyProduct}
  \mc{D}\bigl\{U_l,\epsilon_{L/R}^{U_lG},\tilde{f}_l\bigr\}\ =\ %
  \mc{C}_{\tilde{f}_n}^{U_n,G}\Bigl(\epsilon_{L}^{U_nG},\epsilon_{R}^{U_nG}\Bigr)
  \cdot\ \ldots\ \cdot%
  \mc{C}_{\tilde{f}_1}^{U_1,G}\Bigl(\epsilon_{L}^{U_1G},\epsilon_{R}^{U_1G}\Bigr)\subset G\ \ ,
\end{equation}
  is invariant under the action of $H$ provided that the embedding maps
  satisfy the relations
\begin{equation}
  \label{eq:AutoRestriction}
  \epsilon_{R}^{U_{l+1}G}\circ\epsilon^{HU_{l+1}}
  \ =\ \epsilon_{L}^{U_lG}\circ\epsilon^{HU_l}\ \ .
\end{equation}
  For later purposes we also have to demand that the embedding indices
  of the left and right embeddings $\epsilon_{L/R}^{U_lG}$
  are identical for fixed subgroup $U_l$. Both conditions are automatically
  satisfied if we restrict to the original setup of section \ref{sc:Lagrange}.
  The elements $x\in\mc{D}\bigl\{U_l,\epsilon_{L/R}^{U_lG},\tilde{f}_l\bigr\}$
  transform under these conditions according to
\begin{equation}
  \label{eq:TrafoNew}
  x\ \mapsto\ \epsilon_{L}^{U_nG}\circ\epsilon^{HU_n}\bigl(h\bigr)\cdot x\cdot\epsilon_{R}^{U_1G}\circ\epsilon^{HU_1}\bigl(h^{-1}\bigr)\ \ .
\end{equation}
  To write down the boundary WZNW functional for D-branes candidates which are
  localized along $\mc{D}\bigl\{U_l,\epsilon_{L/R}^{U_lG},\tilde{f}_l\bigr\}$
  we have to generalize the definition \eqref{eq:BoundaryForm} to
\begin{equation*}
  \omega_{\mc{D}}(c_l)
  \ =\ \tr_R\Bigl\{\epsilon_L^{U_lG}\bigl(s_l^{-1}ds_l\bigr)\,\tilde{f}_l\,\epsilon_R^{U_lG}\bigl(s_l^{-1}ds_l\bigr)\,\tilde{f}_l^{-1}\Bigr\}\ \ .
\end{equation*}
\medskip

  In this appendix we will present the computational details that
  a) the boundary WZNW functional \eqref{eq:BoundaryAction} is invariant
  under the infinitesimal action (\ref{eq:Trafo},\,\ref{eq:TrafoNew}) of
  $h=1+i\omega\in H$ on the boundary
  and b) that it is well-defined with respect to infinitesimal deformations
  of the disc $D$. We will, however, not be concerned with global issues
  which give rise to quantization of generalized twisted conjugacy classes
  and branching selection rules. These global topological properties may
  possibly lead to severe restrictions which prohibit certain subsets
  of the form \eqref{eq:ConjugacyProduct}. For example we would not know
  how to model maximally symmetric D-branes which are localized
  along the product of two conjugacy classes
  $\mc{C}_{f_1}^G\cdot\mc{C}_{f_2}^G$ in the algebraic description. This
  example suggests that there also might be problems with the conformal
  invariance of our boundary WZNW functional \eqref{eq:BoundaryAction}
  on the quantum level as there is no algebraic description
  corresponding to this more general setting. These questions have to be
  addressed in future work.
\medskip

  Let us start with item a), i.e.\ the invariance of the action functional
  \eqref{eq:BoundaryAction}
  under transformations of the form (\ref{eq:Trafo},\,\ref{eq:TrafoNew})
  on the boundary. Referring to the discussion in section \ref{sc:Lagrange}
  this amounts to a proof of the relation
  $\delta\omega^{\text{WZ}}\bigr|_{D}=d\delta\omega_{\mc{D}}$.
  The elements of the twisted conjugacy classes transform according to
\begin{equation*}
  \delta c_l
  \ =\ i\,\omega_L^{(l)}\,c_l\ -\ i\,c_l\,\omega_R^{(l)}\ \ ,
\end{equation*}
  where we introduced the short hand notations
  $\omega_{L/R}^{(l)}\ =\ \epsilon_{L/R}^{U_lG}\circ\epsilon^{HU_l}(\omega)$.
  The condition \eqref{eq:AutoRestriction} of mutual consistency of the
  embedding maps obviously translates into the relation
  $\omega_R^{(l+1)}=\omega_L^{(l)}$.
  Supplied with this information it is now very easy to calculate
  all variations
\begin{equation*}
  \delta(c_lc_{l-1}\cdots c_{k+1}c_k)\ =\ i\,\omega_{L}^{(l)}\,c_lc_{l-1}\cdots c_{k+1}c_k\ +\ i\,c_lc_{l-1}\cdots c_{k+1}c_k\,\omega_R^{(k)}\ \ .
\end{equation*}
  Similar relations hold for the inverse
  $\delta c_l^{-1}=i\omega_R^{(l)}c_l^{-1}-ic_l^{-1}\omega_L^{(l)}$
  and for chains of the form $c_k^{-1}c_{k+1}^{-1}\cdots c_{l-1}^{-1}c_l^{-1}$.
  Finally, we also need to know the variation
\begin{equation*}
  \delta dc_l
  \ =\ d\Bigl( i\,\omega_L^{(l)}\,c_l\ -\ i\,c_l\,\omega_R^{(l)}\Bigr)
  \ =\ i\,d\omega_L^{(l)}\,c_l\ -\ i\,c_l\,d\omega_R^{(l)}
     +i\,\omega_L^{(l)}\,dc_l\ -\ i\,dc_l\,\omega_R^{(l)}\ \ .
\end{equation*}
  Due to these relations the transformation properties of
  $\omega_{\mc{D}}(c_k,\cdots,c_l)$ may easily be calculated. It turns out
  that all terms involving $\omega$ cancel each other. Only
  four terms involving $d\omega$ survive. We summarize this result
  in
\begin{eqnarray*}
  \delta\omega_{\mc{D}}(c_k,\cdots,c_l)
  &=&-i\,\tr\bigl\{c_k^{-1}\cdots c_l^{-1}\,d\omega_L^{(l)}\,c_l\cdots c_{k+1}dc_k\bigr\}\\
   &&+i\,\tr\bigl\{c_k^{-1}\cdots c_{l-1}^{-1}\,d\omega_L^{(l-1)}\,c_{l-1}\cdots c_{k+1}dc_k\bigr\}\\
   &&-i\,\tr\bigl\{c_{k+1}^{-1}\cdots c_l^{-1}dc_lc_{l-1}\cdots c_{k+1}\,d\omega_R^{(k+1)}\,\bigr\}\\
   &&+i\,\tr\bigl\{c_k^{-1}\cdots c_l^{-1}dc_lc_{l-1}\cdots c_k\,d\omega_R^{(k)}\bigr\}\ \ .
\end{eqnarray*}
  When evaluating this expression special care has to be taken if $l=k+1$.
  In this case no factors $c_{l-1}\cdots c_{k+1}$ appear between the
  differentials in lines two and three.
\medskip

  Due to its different structure the variation of
  $\omega_{\mc{D}}(c_l)$ has to be treated separately. In this case we obtain
\begin{equation*}
  \delta\omega_{\mc{D}}(c_l)
  \ = \ -i\,\tr\bigl\{d\omega_L^{(l)}dc_lc_l^{-1}+d\omega_R^{(l)}c_l^{-1}dc_l\bigr\}\ \ .
\end{equation*}
  During the calculation we made use of
\begin{eqnarray*}
  c_l^{-1}dc_l
  &=&\epsilon_R^{U_lG}\bigl(s_l\bigr)\,\tilde{f}_l^{-1}\,\epsilon_L^{U_lG}\bigl(s_l^{-1}ds_l\bigr)\,\tilde{f}_l\,\epsilon_R^{U_lG}\bigl(s_l^{-1}\bigr)
     -\epsilon_R^{U_lG}\bigl(ds_ls_l^{-1}\bigr)\\
  dc_lc_l^{-1}
  &=&\epsilon_L^{U_lG}\bigl(ds_ls_l^{-1}\bigr)
     -\epsilon_L^{U_lG}\bigl(s_l\bigr)\,\tilde{f}_l\,\epsilon_R^{U_lG}\bigl(s_l^{-1}ds_l\bigr)\,\tilde{f}_l^{-1}\,\epsilon_L^{U_lG}\bigl(s_l^{-1}\bigr)\ \ .
\end{eqnarray*}
  Indeed, these two relations imply
\begin{equation*}
  i\,\tr\bigl\{d\omega_L^{(l)}dc_lc_l^{-1}+d\omega_R^{(l)}c_l^{-1}dc_l\bigr\}
  =-\delta\omega_{\mc{D}}(c_l)
       +\,i\,\tr\Bigl\{d\omega_L^{(l)}\epsilon_L^{U_lG}\bigl(ds_ls_l^{-1}\bigr)-d\omega_R^{(l)}\epsilon_R^{U_lG}\bigl(ds_ls_l^{-1}\bigr)\Bigr\}.
\end{equation*}
  If we rewrite the last term according to
\begin{equation*}
  \tr\Bigl\{\epsilon_L^{U_lG}\bigl(\epsilon^{HU_l}(d\omega)ds_ls_l^{-1}\bigr)
   -\epsilon_R^{U_lG}\bigl(\epsilon^{HU_l}(d\omega)ds_ls_l^{-1}\bigr)\Bigr\}
\end{equation*}
  we see that it vanishes provided the two embeddings $\epsilon_{L/R}^{U_lG}$
  have the same embedding index.
\medskip

  Summing up all contributions and remembering that the variation of
  $\omega_{\mc{D}}(c_k,c_{k+1})$ shows some subtleties we obtain
\begin{equation*}
  \delta\omega_{\mc{D}}
  \ =\ -i\,\sum_{l=1}^n\tr\bigl\{d\omega_L^{(n)}c_n\cdots c_{l+1}dc_lc_l^{-1}\cdots c_n^{-1}+d\omega_R^{(1)}c_1^{-1}\cdots c_l^{-1}dc_lc_{l-1}\cdots c_1\bigr\}\ \ .
\end{equation*}
  During the calculation we made use of several cancellations.
  Finally, we have to compare this expression with the variation of
  the Wess-Zumino term. A careful calculation gives
\begin{equation*}
  \delta \omega^{\text{WZ}}
  \ =\ -i\,d\,\tr\Bigl\{d\omega_L^{(n)}dgg^{-1}+d\omega_R^{(1)}g^{-1}dg\Bigr\}\ \ .
\end{equation*}
  This may easily be evaluated using the relations
\begin{equation*}
  g^{-1}dg
  \ =\ \sum_{l=1}^nc_1^{-1}\cdots c_l^{-1}dc_lc_{l-1}\cdots c_1
  \ \ ,\qquad\qquad
  dgg^{-1}
  \ =\ \sum_{l=1}^n c_n\cdots c_{l+1}dc_lc_l^{-1}\cdots c_n^{-1}\ \ .
\end{equation*}
  The variation then reads
\begin{equation*}
  \delta \omega^{\text{WZ}}
  \ =\ -i\,\sum_{l=1}^nd\,\tr\Bigl\{d\omega_L^{(n)}c_n\cdots c_{l+1}dc_lc_l^{-1}\cdots c_n^{-1}+d\omega_R^{(1)}c_1^{-1}\cdots c_l^{-1}dc_lc_{l-1}\cdots c_1\Bigr\}\ \ .
\end{equation*}
  Obviously, the contributions from $\delta\omega^{\text{WZ}}$ and
  $\delta\omega_{\mc{D}}$ cancel each other exactly. This proves that the
  product of generalized twisted conjugacy classes is indeed a valid candidate
  for the geometry of D-branes which preserve an action of the group $H$.
\medskip
    
  Now we are able to address item b), i.e.\ the invariance of the action
  functional \eqref{eq:BoundaryAction} under infinitesimal deformations
  of the disc $D$. It is sufficient to proof the relation
  $d\omega_{\mc{D}}=\omega^{\text{WZ}}\bigl|_D$.
  The calculation turns out to be very involved if one tries to perform
  it directly. It is convenient to use an induction argument instead, i.e.\ %
  we supply the boundary two-form with an additional label $n$ and write
  $\omega_{\mc{D}}(n)$. The number $n$ ist just the number of generalized
  twisted conjugacy classes appearing in eq.\ \eqref{eq:ConjugacyProduct}.
  For $n=1$ we have $\omega_{\mc{D}}(1)=\omega_{\mc{D}}(c_1)$. Let us thus
  first determine
\begin{eqnarray*}
  d\omega_{\mc{D}}(c_l)
  &=&-\tr\Bigl\{\epsilon_L^{U_lG}\bigl(s_l^{-1}ds_ls_l^{-1}ds_l\bigr)\,\tilde{f}_l\,\epsilon_R^{U_lG}\bigl(s_l^{-1}ds_l\bigr)\,\tilde{f}_l^{-1}\Bigr\}\\
   &&+\,\tr\Bigl\{\epsilon_L^{U_lG}\bigl(s_l^{-1}ds_l\bigr)\,\tilde{f}_l\,\epsilon_R^{U_lG}\bigl(s_l^{-1}ds_ls_l^{-1}ds_l\bigr)\,\tilde{f}_l^{-1}\Bigr\}\ \ .
\end{eqnarray*}
  On the other hand we have
\begin{eqnarray*}
  \omega^{\text{WZ}}(c_l)
  &=&\frac{1}{3}\,\tr\Bigl\{\Bigl(\epsilon_R^{U_lG}\bigl(s_l\bigr)\,\tilde{f}_l^{-1}\,\epsilon_L^{U_lG}\bigl(s_l^{-1}ds_l\bigr)\,\tilde{f}_l\,\epsilon_R^{U_lG}\bigl(s_l^{-1}\bigr)
     -\epsilon_R^{U_lG}\bigl(ds_ls_l^{-1}\bigr)\Bigr)^3\Bigr\}\\
  &=&\frac{1}{3}\,\tr\Bigl\{\epsilon_L^{U_lG}\bigl((s_l^{-1}ds_l)^3\bigr)
     -\epsilon_R^{U_lG}\bigl((ds_ls_l^{-1})^3\bigr)\Bigr\}\\
   &&-\,\tr\Bigl\{\tilde{f}_l^{-1}\,\epsilon_L^{U_lG}\bigl(s_l^{-1}ds_ls_l^{-1}ds_l\bigr)\,\tilde{f}_l\,\epsilon_R^{U_lG}\bigl(s_l^{-1}ds_l\bigr)\Bigr\}\\
   &&+\,\tr\Bigl\{\,\tilde{f}_l^{-1}\,\epsilon_L^{U_lG}\bigl(s_l^{-1}ds_l\bigr)\,\tilde{f}_l\,\epsilon_R^{U_lG}\bigl(s_l^{-1}ds_ls_l^{-1}ds_l\bigr)\Bigr\}\ \ .
\end{eqnarray*}
  The first two terms vanish as the two embeddings by assumption have
  the same embedding index. By specializing to $l=1$ we have proven
  $d\omega_{\mc{D}}(n)=\omega^{\text{WZ}}\bigl|_D$ for $n=1$.
\medskip

  Let us now turn to the case $n>1$. It is convenient to introduce the
  notation $g_n=c_n\cdots c_1=c_n\,g_{n-1}$. In addition we also need
  the recursion property
  $\omega_{\mc{D}}(n)=\omega_{\mc{D}}(n-1)+\sum_{l=1}^n\omega_{\mc{D}}(c_l,\cdots,c_n)$.
  Using the representation above we easily obtain
  $g_n^{-1}dg_n=g_{n-1}^{-1}c_n^{-1}dc_ng_{n-1}+g_{n-1}^{-1}dg_{n-1}$.
  We are thus able to calculate
\begin{eqnarray*}
  \omega^{\text{WZ}}(g_n)
  &=&\omega^{\text{WZ}}(g_{n-1})\ +\ \omega^{\text{WZ}}(c_l)\\
   &&+\,\tr\bigl\{c_n^{-1}dc_nc_n^{-1}dc_ndg_{n-1}g_{n-1}^{-1}
     \ +\ c_n^{-1}dc_ndg_{n-1}g_{n-1}^{-1}dg_{n-1}g_{n-1}^{-1}\bigr\}\ \ .
\end{eqnarray*}
  By induction we have $d\omega_{\mc{D}}(n-1)=\omega^{\text{WZ}}(g_{n-1})$. We
  also proved already that $d\omega(c_n)=\omega^{\text{WZ}}(c_n)$. It thus
  remains to check whether
\begin{equation}
  \label{eq:aux1}
  \sum_{l=1}^{n-1}\:d\omega_{\mc{D}}(c_l,\cdots,c_n)
  \ =\ \,\tr\bigl\{c_n^{-1}dc_nc_n^{-1}dc_ndg_{n-1}g_{n-1}^{-1}\:
       +\:c_n^{-1}dc_ndg_{n-1}g_{n-1}^{-1}dg_{n-1}g_{n-1}^{-1}\bigr\}\ \ .
\end{equation}
  Indeed, for the left hand side we have
\begin{equation}
  \label{eq:aux4}
  \begin{split}
  \sum_{l=1}^{n-1}\:d\omega_{\mc{D}}(c_l,\cdots,c_n)
  &=\sum_{l=1}^{n-1}\sum_{k=l}^{n-1}\tr\bigl\{c_l^{-1}\cdots c_k^{-1}dc_kc_k^{-1}\cdots c_n^{-1}dc_nc_{n-1}\cdots c_{l+1}dc_l\bigr\}\\
   &+\ \sum_{l=1}^{n-1}\tr\bigl\{c_l^{-1}\cdots c_n^{-1}dc_nc_n^{-1}dc_nc_{n-1}\cdots c_{l+1}dc_l\bigr\}\\
   &+\ \sum_{l=1}^{n-1}\sum_{k=l+1}^{n-1}\tr\bigl\{c_l^{-1}\cdots c_n^{-1}dc_nc_{n-1}\cdots c_{k+1}dc_kc_{k-1}\cdots c_{l+1}dc_l\bigr\}\ \ .
  \end{split}
\end{equation}
  To evaluate the right hand side of eq.\ \eqref{eq:aux1}
  we use the explicit form of $g_{n-1}$ as a product of $c$'s
  and write
\begin{equation*}
  \begin{split}
  g_{n-1}^{-1}dg_{n-1}
  &\ =\ \sum_{l=1}^{n-1}c_1^{-1}\cdots c_l^{-1}dc_lc_{l-1}\cdots c_1\\
  dg_{n-1}g_{n-1}^{-1}
  &\ =\ \sum_{l=1}^{n-1} c_{n-1}\cdots c_{l+1}dc_lc_l^{-1}\cdots c_{n-1}^{-1}\ \ .
  \end{split}
\end{equation*}
  Taking the square of the last expression we arrive at
\begin{eqnarray*}
  dg_{n-1}g_{n-1}^{-1}dg_{n-1}g_{n-1}^{-1}
  &=&\sum_{l=1}^{n-1}\sum_{k=1}^{n-1} c_{n-1}\cdots c_{l+1}dc_lc_l^{-1}\cdots c_{n-1}^{-1}c_{n-1}\cdots c_{k+1}dc_kc_k^{-1}\cdots c_{n-1}^{-1}\\
  &=&\sum_{l=1}^{n-1}\sum_{k=l}^{n-1} c_{n-1}\cdots c_{l+1}dc_lc_l^{-1}\cdots c_k^{-1}dc_kc_k^{-1}\cdots c_{n-1}^{-1}\\
   &&+\sum_{l=1}^{n-1}\sum_{k=l+1}^{n-1} c_{n-1}\cdots c_{k+1}dc_kc_{k-1}\cdots c_{l+1}dc_lc_l^{-1}\cdots c_{n-1}^{-1}\ \ .
\end{eqnarray*}
  Plugging all this into the right hand side of eq.\ \eqref{eq:aux1}
  we finally arrive at
\begin{eqnarray*}
  &&\tr\bigl\{c_n^{-1}dc_nc_n^{-1}dc_ndg_{n-1}g_{n-1}^{-1}\:
       +\:c_n^{-1}dc_ndg_{n-1}g_{n-1}^{-1}dg_{n-1}g_{n-1}^{-1}\bigr\}\\
  &=&\sum_{l=1}^{n-1}\:\tr\Bigl\{c_n^{-1}dc_nc_n^{-1}dc_nc_{n-1}\cdots c_{l+1}dc_lc_l^{-1}\cdots c_{n-1}^{-1}\Bigr\}\\
   &&+\sum_{l=1}^{n-1}\sum_{k=l}^{n-1}\:\tr\Bigl\{c_n^{-1}dc_nc_{n-1}\cdots c_{l+1}dc_lc_l^{-1}\cdots c_k^{-1}dc_kc_k^{-1}\cdots c_{n-1}^{-1}\Bigr\}\\
   &&+\sum_{l=1}^{n-1}\sum_{k=l+1}^{n-1}\:\tr\Bigl\{c_n^{-1}dc_nc_{n-1}\cdots c_{k+1}dc_kc_{k-1}\cdots c_{l+1}dc_lc_l^{-1}\cdots c_{n-1}^{-1}\Bigr\}\ \ .
\end{eqnarray*}
  This expression coincides with the expression \eqref{eq:aux4}.
  We have thus completed our induction argument.

\end{appendix}

\providecommand{\href}[2]{#2}\begingroup\raggedright\endgroup

%\bibliographystyle{$HOME/texstyles/JHEP-2}
%\bibliography{$HOME/bibliography/bibliography}

\end{document}

%% file: SLCoord.pstex_t
\begin{picture}(0,0)%
\epsfig{file=SLCoord.pstex}%
\end{picture}%
\setlength{\unitlength}{4144sp}%
\begingroup\makeatletter\ifx\SetFigFont\undefined%
\gdef\SetFigFont#1#2#3#4#5{%
  \reset@font\fontsize{#1}{#2pt}%
  \fontfamily{#3}\fontseries{#4}\fontshape{#5}%
  \selectfont}%
\fi\endgroup%
\begin{picture}(1284,2174)(1159,-1598)
\put(1891,344){\makebox(0,0)[lb]{\smash{\SetFigFont{10}{12.0}{\familydefault}{\mddefault}{\updefault}$\tau$}}}
\put(2116,-601){\makebox(0,0)[b]{\smash{\SetFigFont{10}{12.0}{\familydefault}{\mddefault}{\updefault}$\theta$}}}
\put(1614,-558){\makebox(0,0)[b]{\smash{\SetFigFont{10}{12.0}{\familydefault}{\mddefault}{\updefault}$\rho$}}}
\end{picture}

%% file: SLTypC1.pstex_t
\begin{picture}(0,0)%
\epsfig{file=SLTypC1.pstex}%
\end{picture}%
\setlength{\unitlength}{4144sp}%
\begingroup\makeatletter\ifx\SetFigFont\undefined%
\gdef\SetFigFont#1#2#3#4#5{%
  \reset@font\fontsize{#1}{#2pt}%
  \fontfamily{#3}\fontseries{#4}\fontshape{#5}%
  \selectfont}%
\fi\endgroup%
\begin{picture}(1284,2174)(1159,-1598)
\put(1891,-556){\makebox(0,0)[lb]{\smash{\SetFigFont{10}{12.0}{\familydefault}{\mddefault}{\updefault}$\tau=0$}}}
\put(1891,344){\makebox(0,0)[lb]{\smash{\SetFigFont{10}{12.0}{\familydefault}{\mddefault}{\updefault}$\tau=\pi$}}}
\end{picture}

%% file: SLTypC2.pstex_t
\begin{picture}(0,0)%
\epsfig{file=SLTypC2.pstex}%
\end{picture}%
\setlength{\unitlength}{4144sp}%
\begingroup\makeatletter\ifx\SetFigFont\undefined%
\gdef\SetFigFont#1#2#3#4#5{%
  \reset@font\fontsize{#1}{#2pt}%
  \fontfamily{#3}\fontseries{#4}\fontshape{#5}%
  \selectfont}%
\fi\endgroup%
\begin{picture}(1284,2174)(1159,-1598)
\end{picture}

%% file: SLTypC3.pstex_t
\begin{picture}(0,0)%
\epsfig{file=SLTypC3.pstex}%
\end{picture}%
\setlength{\unitlength}{4144sp}%
\begingroup\makeatletter\ifx\SetFigFont\undefined%
\gdef\SetFigFont#1#2#3#4#5{%
  \reset@font\fontsize{#1}{#2pt}%
  \fontfamily{#3}\fontseries{#4}\fontshape{#5}%
  \selectfont}%
\fi\endgroup%
\begin{picture}(1284,2174)(1159,-1598)
\end{picture}

%% file: SLTypT1.pstex_t
\begin{picture}(0,0)%
\epsfig{file=SLTypT1.pstex}%
\end{picture}%
\setlength{\unitlength}{4144sp}%
\begingroup\makeatletter\ifx\SetFigFont\undefined%
\gdef\SetFigFont#1#2#3#4#5{%
  \reset@font\fontsize{#1}{#2pt}%
  \fontfamily{#3}\fontseries{#4}\fontshape{#5}%
  \selectfont}%
\fi\endgroup%
\begin{picture}(1284,2184)(1159,-1603)
\end{picture}

%% file: SLTypS1.pstex_t
\begin{picture}(0,0)%
\epsfig{file=SLTypS1.pstex}%
\end{picture}%
\setlength{\unitlength}{4144sp}%
\begingroup\makeatletter\ifx\SetFigFont\undefined%
\gdef\SetFigFont#1#2#3#4#5{%
  \reset@font\fontsize{#1}{#2pt}%
  \fontfamily{#3}\fontseries{#4}\fontshape{#5}%
  \selectfont}%
\fi\endgroup%
\begin{picture}(1284,2174)(1159,-1598)
\end{picture}

%% file: SLTypS2.pstex_t
\begin{picture}(0,0)%
\epsfig{file=SLTypS2.pstex}%
\end{picture}%
\setlength{\unitlength}{4144sp}%
\begingroup\makeatletter\ifx\SetFigFont\undefined%
\gdef\SetFigFont#1#2#3#4#5{%
  \reset@font\fontsize{#1}{#2pt}%
  \fontfamily{#3}\fontseries{#4}\fontshape{#5}%
  \selectfont}%
\fi\endgroup%
\begin{picture}(1284,2174)(1159,-1598)
\end{picture}

%% file: SLTypS3.pstex_t
\begin{picture}(0,0)%
\epsfig{file=SLTypS3.pstex}%
\end{picture}%
\setlength{\unitlength}{4144sp}%
\begingroup\makeatletter\ifx\SetFigFont\undefined%
\gdef\SetFigFont#1#2#3#4#5{%
  \reset@font\fontsize{#1}{#2pt}%
  \fontfamily{#3}\fontseries{#4}\fontshape{#5}%
  \selectfont}%
\fi\endgroup%
\begin{picture}(1284,2174)(1159,-1598)
\end{picture}

%% file: SLTypS4.pstex_t
\begin{picture}(0,0)%
\epsfig{file=SLTypS4.pstex}%
\end{picture}%
\setlength{\unitlength}{4144sp}%
\begingroup\makeatletter\ifx\SetFigFont\undefined%
\gdef\SetFigFont#1#2#3#4#5{%
  \reset@font\fontsize{#1}{#2pt}%
  \fontfamily{#3}\fontseries{#4}\fontshape{#5}%
  \selectfont}%
\fi\endgroup%
\begin{picture}(1284,2174)(1159,-1598)
\end{picture}

%% file: SUTypC1.pstex_t
\begin{picture}(0,0)%
\epsfig{file=SUTypC1.pstex}%
\end{picture}%
\setlength{\unitlength}{4144sp}%
\begingroup\makeatletter\ifx\SetFigFont\undefined%
\gdef\SetFigFont#1#2#3#4#5{%
  \reset@font\fontsize{#1}{#2pt}%
  \fontfamily{#3}\fontseries{#4}\fontshape{#5}%
  \selectfont}%
\fi\endgroup%
\begin{picture}(1816,1886)(893,-1454)
\put(1906,-284){\makebox(0,0)[b]{\smash{\SetFigFont{10}{12.0}{\familydefault}{\mddefault}{\updefault}$\frac{\pi\mu}{k}$}}}
\end{picture}

%% file: SURot.pstex_t
\begin{picture}(0,0)%
\epsfig{file=SURot.pstex}%
\end{picture}%
\setlength{\unitlength}{4144sp}%
\begingroup\makeatletter\ifx\SetFigFont\undefined%
\gdef\SetFigFont#1#2#3#4#5{%
  \reset@font\fontsize{#1}{#2pt}%
  \fontfamily{#3}\fontseries{#4}\fontshape{#5}%
  \selectfont}%
\fi\endgroup%
\begin{picture}(1816,1819)(893,-1418)
\end{picture}

%% file: SUTypS1.pstex_t
\begin{picture}(0,0)%
\epsfig{file=SUTypS1.pstex}%
\end{picture}%
\setlength{\unitlength}{4144sp}%
\begingroup\makeatletter\ifx\SetFigFont\undefined%
\gdef\SetFigFont#1#2#3#4#5{%
  \reset@font\fontsize{#1}{#2pt}%
  \fontfamily{#3}\fontseries{#4}\fontshape{#5}%
  \selectfont}%
\fi\endgroup%
\begin{picture}(1816,1814)(803,-1418)
\end{picture}